\pdfoutput=1
\RequirePackage{ifpdf}
\ifpdf % We are running pdfTeX in pdf mode
\documentclass[pdftex]{sigma}
\else
\documentclass{sigma}
\fi

\usepackage{eepic,texdraw,setspace,tikz}
\usetikzlibrary{arrows, decorations.pathmorphing, backgrounds, positioning, fit, petri, automata}

\numberwithin{equation}{section}

\usepackage{subfigure}

{ \theoremstyle{definition}
\newtheorem{rem}{Remark}[section]
}
\newtheorem{Prop}[rem]{Proposition}

\newcommand{\bp}{\begin{Prop}}
\newcommand{\ep}{\end{Prop}}
\newcommand{\br}{\begin{rem}}
\newcommand{\er}{\end{rem}}

\newcommand{\pa}{\partial}

\begin{document}

\allowdisplaybreaks

\newcommand{\arXivNumber}{1708.07024}

\renewcommand{\PaperNumber}{022}

\FirstPageHeading

\ShortArticleName{Poisson Algebras and 3D Superintegrable Hamiltonian Systems}

\ArticleName{Poisson Algebras and 3D Superintegrable\\ Hamiltonian Systems}

\Author{Allan P.~FORDY~$^\dag$ and Qing HUANG~$^\ddag$}

\AuthorNameForHeading{A.P.~Fordy and Q.~Huang}

\Address{$^\dag$~School of Mathematics, University of Leeds, Leeds LS2 9JT, UK}
\EmailD{\href{mailto:A.P.Fordy@leeds.ac.uk}{A.P.Fordy@leeds.ac.uk}}

\Address{$^\ddag$~School of Mathematics, Northwest University, Xi'an 710069, People's Republic of China}
\EmailD{\href{mailto:hqing@nwu.edu.cn}{hqing@nwu.edu.cn}}

\ArticleDates{Received August 24, 2017, in final form March 06, 2018; Published online March 16, 2018}

\Abstract{Using a Poisson bracket representation, in 3D, of the Lie algebra $\mathfrak{sl}(2)$, we first use highest weight representations to embed this into larger Lie algebras. These are then interpreted as symmetry and conformal symmetry algebras of the ``kinetic energy'', related to the quadratic Casimir function. We then consider the potentials which can be added, whilst remaining integrable, leading to families of separable systems, depending upon arbitrary functions of a single variable. Adding further integrals, in the superintegrable case, restricts these functions to specific forms, depending upon a finite number of arbitrary parameters. The Poisson algebras of these superintegrable systems are studied. The automorphisms of the symmetry algebra of the kinetic energy are extended to the full Poisson algebra, enabling us to build the full set of Poisson relations.}


\Classification{17B63; 37J15; 37J35; 70G45; 70G65; 70H06}

\section{Introduction}

This paper is in two parts. Sections \ref{Sect:basic}--\ref{Sect:conformal} are mainly algebraic, building Lie algebras with a~given copy of $\mathfrak{sl}(2)$ as a subalgebra. Since the second part of the paper (Sections \ref{Sect:separable} and~\ref{Sect:super}) is about completely integrable Hamiltonian systems (and their {\em super-integrable} restrictions), the Lie algebraic part is presented in a Poisson bracket representation (with~$3$ degrees of freedom), so we are constructing Poisson algebras with linear relations. Our emphasis is on building a~Poisson algebra with a desired Lie algebraic structure.

We extend the $3$-dimensional algebra $\mathfrak{sl}(2)$ to $6$- and $10$-dimensional algebras. The quadratic Casimir of the $6$-dimensional algebra can be written in the form (with $n=3$)
\begin{gather*}
H_0 = \frac{1}{2} \sum_{i,j=1}^n g^{ij}({\boldsymbol{q}}) p_ip_j.
\end{gather*}
When the matrix of coefficients $g^{ij}$ is nonsingular, it may be considered as the inverse of a~metric tensor $g_{ij}$ and the function $H_0$ represents the kinetic energy of a freely moving particle on the corresponding manifold (geodesic motion). For a metric with isometries, the infinitesimal generators (Killing vectors) correspond to functions which are {\it linear} in momenta and which Poisson commute with the kinetic energy $H_0$ (the corresponding Noether integrals). When the space is either flat or constant curvature, it possesses the maximal group of isometries, which is of dimension~$\frac{1}{2}n(n+1)$. In this case, $H_0$ is actually the second order {\em Casimir} function of the symmetry algebra (see~\cite{74-7}). This is exactly the case we have, with $n=3$ and a~$6$-dimensional isometry algebra. The maximal number of conformal symmetries (including isometries as a~subalgebra) is of dimension $\frac{1}{2}(n+1)(n+2)=10$, when $n=3$. Our $10$-dimensional extensions are just the corresponding conformal algebras. When $g^{ij}$ is {\em singular}, the Poisson algebras have the same structure, but without the geometric interpretation.

Our main application of the algebraic structures we construct is to build some superintegrable systems with nontrivial, nonlinear Poisson algebras, which generalise the {\em Lie algebraic} Poisson algebras of Sections \ref{Sect:basic}--\ref{Sect:conformal}. Below we give a brief reminder of the meaning of complete and super-integrability.

A Hamiltonian system of $n$ degrees of freedom, Hamiltonian~$H$, is said to be {\em completely integrable in the Liouville sense} if we have $n$ independent functions~$I_n$, which are {\em in involution} (mutually Poisson commuting), with $H$ being a function of these and typically just one of them. Whilst~$n$ is the maximal number of independent functions which can be {\em in involution}, it is possible to have further integrals of the Hamiltonian~$H$, which necessarily generate a non-Abelian algebra of integrals of~$H$. The maximal number of additional {\em independent} integrals is $n-1$, since the ``level surface'' of $2n-1$ integrals (meaning the intersection of individual level surfaces) is just the (unparameterised) integral curve. Well known elementary examples are the isotropic harmonic oscillator, the Kepler system and the Calogero--Moser system. The quadratures of complete integrability are often achieved through the separation of variables of the Hamilton--Jacobi equation. The solution of a~maximally super-integrable system can also be calculated purely algebraically (albeit implicitly), requiring just the solution of the equations $I_k=c_k$, $k=1,\dots ,2n-1$. Maximally superintegrable systems have a number of interesting properties: they can be separable in more than one coordinate system; all bounded orbits are closed; they give rise to interesting Poisson algebras with polynomial Poisson relations. The idea can be extended to {\em quantum integrable systems}, with first integrals replaced by commuting differential operators. For some examples of superintegrable quantum systems it is possible to use the additional commuting operators to build sequences of eigenfunctions~\cite{f07-1,f13-1}. There is a~large literature on the classification and analysis of superintegrable systems (see the review~\cite{13-2}) and they naturally occur in many applications in physics (additional integrals being referred to as ``hidden symmetries''~\cite{14-2}).

Clearly our geodesic flow, with Hamiltonian $H_0$ is super-integrable. There are $6$ Noether integrals, but only $5$ are functionally independent, since there is a quadratic constraint on the $6$-dimensional algebra (see equation (\ref{zero-cas})). Furthermore, each element of the algebra commutes with at least one other element (see Table~\ref{Tab:6Symm_alg}), so the Hamiltonian~$H_0$ belongs to several involutive triples, each of which renders it completely integrable.

In Section~\ref{Sect:separable} we show how to use the symmetry algebra of the kinetic energy $H_0$ to build quadratic (in momenta) integrals, and to add potential functions to build completely integrable systems, which are, in fact, separable. Explicitly, we extend the Hamiltonian functions~$H_0$ to include a potential function:
\begin{gather*}
H = H_0 + h({\boldsymbol{q}}),
\end{gather*}
and use the symmetry algebra of $H_0$ to construct two functions
\begin{gather*}
F_i = K_i + g_i({\boldsymbol{q}}), \qquad\mbox{with}\qquad \{H,F_i\}=\{F_1,F_2\}=0,\qquad i=1,2,
\end{gather*}
where the functions $K_i$ are {\em quadratic forms} of the Noether constants of~$H_0$. For some particular examples, we derive the explicit form of the functions $h$, $g_i$, which depend upon three arbitrary functions of a single variable (the three separation variables). Some examples are related through a Lie algebra automorphism, a property that would not be easy to see without the relation to the symmetry algebra of~$H_0$.

In Section \ref{Sect:super} we consider the superintegrable restrictions of these separable systems, by adding two further integrals. These now typically depend upon a small number of {\em arbitrary parameters} instead of arbitrary functions. Whilst some of these functions still Poisson commute, not all of them can and the non-zero Poisson relations are no longer {\em linear}, but polynomial. In $3$ degrees of freedom, these are considerably more complicated than in the $2$ degrees of freedom case of \cite{06-7,f07-1,f17-1}, and it's not clear that we can always close the algebra in a finite way. However, since our kinetic energy has a $6$-dimensional symmetry algebra with automorphisms which can be realised as canonical transformations, these act on the nonlinear Poisson algebra of our superintegrable systems, enabling us to obtain the full set of Poisson relations.

\section{The basic setup}\label{Sect:basic}

We consider a $6$-dimensional space $M$, with (local) canonical coordinates $q_i$, $p_i$, $i=1,2,3$, satisfying the standard canonical relations
$\{q_i,q_j\}=\{p_i,p_j\}=0$, $\{q_i,p_j\}=\delta_{ij}$, for all $i,j=1,2,3$.

Recall that if $f, g$ are any functions on $M$, then the Hamiltonian vector field of $f$ is
\begin{gather*}
X_f=\sum_{i=1}^3 \left(\{q_i,f\}\pa_{q_i}+\{p_i,f\}\pa_{p_i}\right) \qquad\mbox{and}\qquad [X_f,X_g]=-X_{\{f,g\}}.
\end{gather*}
Functions which are {\it linear} in momenta define vector fields on configuration space, with coordinates $(q_1,q_2,q_3)$. For any function on configuration space, $f(q_1,q_2,q_3)$, we have
\begin{gather}\label{3dVFs}
h({\boldsymbol{q}},{\bf p}) = \sum_{i=1}^3 a_i({\boldsymbol{q}})p_i \quad\Rightarrow\quad \{f,h\} = \sum_{i=1}^3 a_i({\boldsymbol{q}}) \frac{\pa f}{\pa q_i}.
\end{gather}
Later, we use this to represent a Killing vector by its Noether constant, which is linear in momenta.

\subsection[The Lie algebra $\mathfrak{g}_1\simeq \mathfrak{sl}(2)$]{The Lie algebra $\boldsymbol{\mathfrak{g}_1\simeq \mathfrak{sl}(2)}$}

We start with a Poisson representation of the Lie algebra $\mathfrak{sl}(2)$,
\begin{subequations}\label{g1}
\begin{gather}
e_1 = p_2, \qquad h_1 = -2(q_1p_1+q_2p_2+q_3p_3), \nonumber\\
 f_1 = -2q_1q_2p_1+\big(q_3^2-q_1^2-q_2^2\big)p_2-2q_2q_3p_3 ,\label{g1-ehf}
\end{gather}
satisfying
\begin{gather}\label{g1-pbs}
\{e_1,h_1\}=2 e_1,\qquad \{f_1,e_1\}=h_1,\qquad \{f_1,h_1\}=-2 f_1.
\end{gather}
\end{subequations}
With this choice, the corresponding Hamiltonian vector fields will satisfy the standard commutation rules for~$\mathfrak{sl}(2)$.

We can calculate the most general function on this phase space which commutes with the whole algebra~$\mathfrak{g}_1$, which is a function of~$3$ variables:
\bp[general invariant of~$\mathfrak{g}_1$]
The most general function, $I_1$, on our phase space, satisfying $\{e_1,I_1\}=\{h_1,I_1\}=\{f_1,I_1\}=0$, is given by
\begin{gather}\label{g1-inv}
I_1 = F(r_0,r_1,r_2),
\end{gather}
with
\begin{gather*} r_0=\frac{q_3}{q_1},\qquad r_1=q_3p_1+q_1p_3,\qquad r_2 = -\big(q_1^2+q_3^2\big)p_1^2+q_1^2p_2^2-2q_1q_3p_1p_3,
\end{gather*}
where $F$ is an arbitrary function of $3$ variables.
\ep

In particular, the Casimir function is given by
\begin{gather}\label{C1}
{\cal C}_1 = e_1f_1+\frac{1}{4} h_1^2 = r_0^2 r_1^2+\big(r_0^2-1\big) r_2.
\end{gather}
\br
In $2$-dimensions, {\it all} invariants would be just functions of the quadratic Casimir, but in this larger space the general invariant includes all the Casimirs of larger algebras containing $\mathfrak{g}_1$ as a~subalgebra.
\er

The most general {\em quadratic} (in momenta) function of the form (\ref{g1-inv}) is given by
\begin{gather}
I_2 = \psi(r_0) r_1^2-\varphi(r_0) r_2\nonumber\\
 \hphantom{I_2}{} = \varphi\left(\frac{q_3}{q_1}\right) q_1^2 \big(p_1^2-p_2^2-p_3^2\big) +\left(\varphi\left(\frac{q_3}{q_1}\right)+\psi\left(\frac{q_3}{q_1}\right)\right)\big(q_3p_1+q_1p_3\big)^2.\label{g1-I2}
\end{gather}
The determinant of the matrix of coefficients, $G^{ij}$, is $\det G = \varphi^2 q_1^6\big(\varphi r_0^2+\big(r_0^2-1\big) \psi\big)$. When this is nonzero, $G$ defines a {\em conformally flat} metric, but the Ricci scalar is generally a complicated differential expression in the functions~$\varphi$ and~$\psi$, even in the diagonal case, for which $\psi=-\varphi$. In the diagonal case there are two interesting cases:
\begin{enumerate}\itemsep=0pt
 \item[1)] $\varphi=(c_1 r_0+c_2)^2$, which is a constant curvature space with $R=6\big(c_1^2-c_2^2\big)$,
 \item[2)] $\varphi=c_1 \big(r_0^2-1\big)$, which is {\it not} a constant curvature space, but does have constant scalar curvature $R=2c_1$.
\end{enumerate}

In Section \ref{conflat-sep} we consider the involutive system
\begin{gather*}
H = H_0 + h,\qquad F_1 = {\cal C}_1 + g_1,\qquad F_2 = K^2 +g_2,
\end{gather*}
where $H_0=I_2$ is {\em conformally flat} (but not constant curvature) and $K$ is some element of~$\mathfrak{g}_1$, to determine separable choices of potential function.

\subsubsection{Lie algebra automorphisms as canonical transformations}\label{automorphisms}

We can build the standard Lie algebra automorphisms of $\mathfrak{sl}(2)$ as canonical transformations. We denote by $\iota_1$ and $\iota_2$, the {\em involutive automorphisms}
\begin{subequations}\label{i12}
\begin{gather}
\iota_1\colon \ (e_1,h_1,f_1) \mapsto (f_1,-h_1,e_1), \label{i1}\\
\iota_2\colon \ (e_1,h_1,f_1) \mapsto (-e_1,h_1,-f_1), \label{i2}
\end{gather}
which can be realised by canonical transformations, generated by
\begin{gather}\label{s1s2}
 S_1 = \frac{q_1P_1-q_2P_2+q_3P_3}{q_1^2-q_2^2-q_3^2}, \qquad S_2 = q_1P_1-q_2P_2+q_3P_3.
\end{gather}
\end{subequations}
Each of the functions $r_0$, $r_1$, $r_2$ of (\ref{g1-inv}) is invariant under these automorphisms.

\subsection[Our choice of $\mathfrak{sl}(2)$]{Our choice of $\boldsymbol{\mathfrak{sl}(2)}$}

The calculations of this paper could be carried out for any choice of representation of $\mathfrak{sl}(2)$. Perhaps the most natural choice would be the {\em linear representation}
\begin{gather}\label{ehf-linear}
e_1=2 (Q_1 P_2+Q_2 P_3),\qquad h_1 = 2 (Q_1 P_1-Q_3 P_3), \qquad f_1 = Q_2 P_1+Q_3 P_2,
\end{gather}
which is related to the representation (\ref{g1-ehf}) through the point transformation
\begin{gather*}
Q_1=\frac{1}{q_1},\qquad Q_2 = \frac{2 q_2}{q_1},\qquad Q_3 = \frac{2\big(q_2^2+q_3^2-q_1^2\big)}{q_1}.
\end{gather*}
However, the first step in our calculation of Section~\ref{hw-g1} is to seek functions which {\em commute} with~$e_1$, so it is natural to transform~$e_1$ to~$p_i$ for some~$i$. The invariants of $e_1$ are $Q_1$ and $2Q_1Q_3-Q_2^2$, and we have $\big\{\frac{Q_2}{2Q_1},e_1\big\}=1$, so we initially choose
\begin{gather*}
q_1 = \rho(Q_1),\qquad q_2 = \frac{Q_2}{2Q_1}, \qquad q_3=\Theta\big(Q_1,2 Q_1Q_3-Q_2^2\big),
\end{gather*}
which imply
\begin{gather*}
 \{q_1,h_1\}=2 Q_1 \rho'(Q_1),\qquad \{q_2,h_1\}=-2q_2, \qquad \{q_3,h_1\}= 2 Q_1 \Theta_1\big(Q_1,2 Q_1Q_3-Q_2^2\big),
\end{gather*}
where $\Theta_1(y_1,y_2)$ is the partial derivative of $\Theta(y_1,y_2)$ with respect to~$y_1$.

If we {\em choose} to have a symmetric formula for~$h_1$, so that $\{q_i,h_1\}=-2 q_i$, then
\begin{gather*}
q_1 = \frac{1}{Q_1},\qquad q_3=\frac{\sigma\big(2 Q_1Q_3-Q_2^2\big)}{Q_1},
\end{gather*}
with $q_3$ defined up to an arbitrary function of {\em one} variable only. The inverse of this transformation is just
\begin{gather*}
Q_1=\frac{1}{q_1},\qquad Q_2 = \frac{2 q_2}{q_1},\qquad Q_3 = \frac{1}{2} q_1 \left(\frac{4 q_2^2}{q_1^2} + \sigma^{-1}\left(\frac{q_3}{q_1}\right)\right),
\end{gather*}
and the corresponding canonical transformation, with (\ref{ehf-linear}), gives
\begin{subequations}
\begin{gather}
e_1=p_2,\qquad h_1 = -2(q_1p_1+q_2p_2+q_3p_3),\nonumber\\
 f_1 = -2 q_1q_2 p_1+\left(\frac{1}{4} \sigma^{-1}\left(\frac{q_3}{q_1}\right) q_1^2-q_2^2\right)p_2-2 q_2q_3 p_3,\label{ehf-sig}
\end{gather}
which is exactly (\ref{g1-ehf}) when $\sigma^{-1}(r_0) = 4\big(r_0^2-1\big)$, so $\sigma(y_2)=\frac{1}{2} \sqrt{y_2+4}$.

In fact, given this choice of $e_1$, $h_1$, the most general form of $f_1$ is determined only up to {\em $3$ arbitrary functions}:
\begin{gather}\label{general-f1}
f_1 = \big(u q_1^2 -2 q_1q_2\big)p_1+\big(v q_1^2-q_2^2\big)p_2+\big(w q_1^2-2 q_2q_3\big)p_3,
\end{gather}
\end{subequations}
$u$, $v$, $w$ being arbitrary functions of $\frac{q_3}{q_1}$. Our transformed elements (\ref{ehf-sig}) just correspond to $u=w=0, v=\frac{1}{4}\sigma^{-1}$.

\br
Lie classified the $2$-dimensional realisations of $\mathfrak{sl}(2,{\mathbb C})$ and $\mathfrak{sl}(2,{\mathbb R})$. There are~$5$ inequivalent realisations of $\mathfrak{sl}(2,{\mathbb R})$ (see \cite[Section~2]{96-6}). No such classification exists for $3$-dimensional realisations, but not all choices of~$u$, $v$, $w$ in~(\ref{general-f1}) lead to equivalent realisations. For example, the determinant of the matrix of coefficients in~(\ref{C1}) (for general~$u$,~$v$,~$w$) vanishes when $q_1 w\big(\frac{q_3}{q_1}\big) - q_3 u\big(\frac{q_3}{q_1}\big)=0$ (as in our case) and this cannot be equivalent to a non-degenerate case.
\er

\subsection[Highest weight representations of $\mathfrak{g}_1$]{Highest weight representations of $\boldsymbol{\mathfrak{g}_1}$}\label{hw-g1}

We now build highest weight representations of $\mathfrak{g}_1$, starting with
\begin{gather*}
Z_1=A(q_1,q_2,q_3)p_1+B(q_1,q_2,q_3)p_2+C(q_1,q_2,q_3)p_3
\end{gather*} and requiring (\ref{rel:z1}) below, which leads immediately to
\begin{subequations}\label{zi}
\begin{gather}\label{def:z1}
 Z_1=q_1^{1-\frac{\lambda}{2}}\left(A\left(\frac{q_3}{q_1}\right)p_1+B\left(\frac{q_3}{q_1}\right)p_2+C\left(\frac{q_3}{q_1}\right)p_3\right),
\end{gather}
satisfying
\begin{gather}\label{rel:z1}
 \{Z_1,e_1\}=0,\qquad \{Z_1,h_1\}=\lambda Z_1.
\end{gather}
Defining
\begin{gather}\label{Zi}
 Z_{i+1}=\{Z_i,f_1\},\qquad i=1,2,\dots ,
\end{gather}
the Poisson relations (\ref{g1-pbs}) then imply
\begin{gather}\label{Zi-pbs}
\{Z_i,h_1\}=(\lambda-2i+2)Z_i \qquad\mbox{and}\qquad \{Z_i,e_1\} = (i-1) (\lambda-i+2) Z_{i-1}.
\end{gather}
\end{subequations}
From this point, $A$, $B$ and $C$ are functions of $r_0=\frac{q_3}{q_1}$.

For general $\lambda$, we have an infinite-dimensional representation, but when $\lambda=2m$ ($m$ a positive integer), it is finite, with dimension $2m+1$. We are particularly interested in the cases $m=0, 1$.
\begin{enumerate}\itemsep=0pt
 \item {\it The case $m=0$:} %\label{case-m=0}
 It is easy to see that the general formula for $Z_2$ is given by
 \begin{gather*}
 Z_2 = q_1^{1-\frac{\lambda}{2}}\big((\lambda q_2 A+2q_1 B)p_1+(\lambda q_2 B+2q_1 A-2q_3 C)p_2+(\lambda q_2 C+2q_3 B)p_3\big),
 \end{gather*}
 so that, for $Z_2=0$, we require $\lambda=0$, $A = \frac{q_3}{q_1} C$ and $B=0$. This leads to
 \begin{gather}\label{1drep}
 Z_1 = C\left(\frac{q_3}{q_1}\right) (q_3p_1+q_1p_3) \qquad\mbox{and}\qquad \{Z_1,e_1\}=\{Z_1,f_1\}=\{Z_1,h_1\}=0.
 \end{gather}
 In the notation of (\ref{g1-inv}), $Z_1=C(r_0) r_1$.

 \item {\it The case $m=1$:} %\label{case-m=1}
For $m\geq 1$ and $\lambda = 2m$, we automatically have $Z_{2m+2}=0$, without any restrictions on the functions $A$, $B$, $C$. When $m=1$, we have
 \begin{gather}
 Z_1 = A p_1+ B p_2+ C p_3,\nonumber\\
 Z_2 = 2(A q_2+ B q_1)p_1+2(A q_1+ B q_2- C q_3)p_2+2(C q_2+ B q_3)p_3, \label{3dZi} \\
 Z_3 = 2\big(A\big(q_1^2+q_2^2+q_3^2\big)+2 B q_1q_2-2 C q_1q_3\big)p_1+2\big(B\big(q_1^2+q_2^2-q_3^2\big)\nonumber\\
\hphantom{Z_3 =}{} +2A q_1q_2-2 C q_2q_3\big)p_2+2\big(C\big(q_2^2-q_1^2-q_3^2\big)+2A q_1q_3+2 B q_2q_3\big)p_3\nonumber.
\end{gather}
The Poisson bracket relations (\ref{Zi}) and (\ref{Zi-pbs}) take the explicit form
\begin{gather}
 \{Z_1,h_1\}=2Z_1,\qquad \{Z_1,f_1\}=Z_2,\qquad \{Z_2,e_1\}=2Z_1,\qquad \{Z_2,f_1\}=Z_3, \nonumber\\
 \label{Zi3pb}
 \{Z_3,e_1\}=2Z_2,\qquad \{Z_3,h_1\}=-2Z_3,\qquad \{Z_1,e_1\}= \{Z_2,h_1\} = \{Z_3,f_1\} = 0.
\end{gather}
\end{enumerate}

\section{Building Poisson algebras}\label{Sect:PoisAlg}

We have a Lie algebra $\mathfrak{g}_1$ and its action (through the Poisson bracket) on the representation space $\{Z_i\}_{i=1}^{2m+1}$. We may consider the {\em linear space} spanned by $\mathfrak{g}_1$ and $\mathfrak{g}_2$:
\begin{gather*}
\mathfrak{g} = \mathfrak{g}_1 + \mathfrak{g}_2, \qquad\mbox{where}\qquad \mathfrak{g}_2 = \{Z_i\}_{i=1}^{2m+1} \qquad\mbox{and}\qquad
 \{\mathfrak{g}_1,\mathfrak{g}_1\}\subset\mathfrak{g}_1,\qquad \{\mathfrak{g}_1,\mathfrak{g}_2\}\subset\mathfrak{g}_2,
\end{gather*}
but need to specify the possible forms of $\{\mathfrak{g}_2,\mathfrak{g}_2\}$ if we wish to consider $\mathfrak{g}$ as a Lie algebra.

Using this approach, we now build $6$-dimensional algebras. The quadratic Casimir function of the algebra $\mathfrak{g}$ defines a matrix, which can be interpreted as an upper-index metric, when it is non-singular. In this case its inverse defines a metric with Killing vectors corresponding to the elements of $\mathfrak{g}$.

The calculation splits into two parts. First of all we use the Jacobi identity to derive abstract relations. Then, in Section \ref{g2g2=g1} we use these relations to restrict the functions in the concrete realisation of (\ref{3dZi}).

In Section \ref{Sect:conformal}, we further extend to $10$-dimensional algebras, which can be interpreted as conformal symmetry algebras of these metrics.

\subsection{6-dimensional extensions}

If we consider $\mathfrak{g}_2$ to have the basis $Z_1$, $Z_2$, $Z_3$, defined by~(\ref{3dZi}), then it follows from the Poisson bracket relations (\ref{Zi3pb}), that
\begin{gather*}
\{Z_1,h_1\}=2 Z_1,\qquad \{Z_2,h_1\}=0, \qquad \{Z_3,h_1\}=-2 Z_3,
\end{gather*}
so, for this $3$-dimensional invariant space we introduce the notation
\begin{gather*}%\label{ehf}
e_2 = Z_1,\qquad h_2 = Z_2,\qquad f_2 = Z_3.
\end{gather*}
For $\mathfrak{g}$ to form a Lie algebra, we must have
\begin{gather*}
\{\mathfrak{g}_2,\mathfrak{g}_2\}\subset \mathfrak{g}_1 + \mathfrak{g}_2,
\end{gather*}
including the special case when $\mathfrak{g}_2$ is {\em Abelian}.

Noting that $\{\{e_2,h_2\},h_1\}= 2 \{e_2,h_2\}$, we have
\begin{gather*}
\{e_2,h_2\} = \alpha e_1 + \beta e_2,\qquad\mbox{for arbitrary constants}\ \alpha, \ \beta.
\end{gather*}
The action of $f_1$ leads to
\begin{gather*}
\{e_2,f_2\} = -\alpha h_1 + \beta h_2,\qquad\mbox{and} \qquad \{h_2,f_2\} = -2\alpha f_1 + \beta f_2.
\end{gather*}
In fact, we may choose $\beta=0$ without loss of generality, as shown by the following:
\bp[$\{\mathfrak{g}_2,\mathfrak{g}_2\}\subset\mathfrak{g}_1$]\label{prop:g2g2}
If the vector space $\mathfrak{g} = \mathfrak{g}_1 + \mathfrak{g}_2$ forms a Lie algebra, satisfying the Poisson bracket relations \eqref{g1-pbs} and \eqref{Zi3pb}, then a basis can be chosen for $\mathfrak{g}_2$, satisfying $\{\mathfrak{g}_2,\mathfrak{g}_2\}\subset\mathfrak{g}_1$. Specifically, there exists a parameter $a$, such that
\begin{gather}\label{e2h2=ae1}
\{e_2,h_2\}=a e_1,\qquad \{e_2,f_2\}=-a h_1,\qquad \{h_2,f_2\}=-2a f_1.
\end{gather}
The possibility of $a=0$ is included, in which case $\mathfrak{g}_2$ forms an Abelian subalgebra.
\ep

\begin{proof}
Defining
\begin{gather*}
\hat e_2 = e_2 + \gamma e_1 \quad\Rightarrow\quad \hat h_2 = h_2 - \gamma h_1 \qquad\mbox{and}\qquad \hat f_2 = f_2 - 2\gamma f_1,
\end{gather*}
for arbitrary parameter $\gamma$, then
\begin{gather*}
\{\hat e_2,\hat h_2\} = \big(\alpha-2 \gamma^2\big) e_1 + (\beta-4\gamma) e_2.
\end{gather*}
Choosing $\gamma=\frac{1}{4} \beta$, we have
\begin{gather*}
\big\{\hat e_2,\hat h_2\big\} = \hat \alpha e_1,\qquad\mbox{where}\qquad \hat \alpha = \alpha -\frac{1}{8} \beta^2.
\end{gather*}
The action of $f_1$ then leads to
\begin{gather*}
\big\{\hat e_2,\hat f_2\big\} = -\hat\alpha h_1 ,\qquad\mbox{and} \qquad \big\{\hat h_2,\hat f_2\big\} = -2\hat\alpha f_1,
\end{gather*}
giving (\ref{e2h2=ae1}) after dropping ``hats'' and setting $\hat\alpha=a$.
\end{proof}

{\bf Casimir functions.}
This $6$-dimensional algebra has a quadratic Casimir function
\begin{gather}\label{Cas12}
 {\cal C}_{12}= 2a \left(e_1f_1+\frac{1}{4} h_1^2\right)+ 2e_2 f_2-h_2^2,
\end{gather}
which will play an important role in what follows.

As an abstract (rank $2$) algebra, there is a second independent (fourth order) Casimir element
\begin{gather*}
{\cal C}_{4} = \big(e_1 f_2+f_2 e_1 +h_1 h_2+h_2 h_1-2 (f_1 e_2+e_2 f_1)\big)^2.
\end{gather*}
In the $6\times 6$ matrix representation (the adjoint representation), this is a multiple of the identity matrix, but in our Poisson representation, it vanishes identically, so the $6$-dimensional Poisson algebra has a quadratic constraint:
\begin{gather}\label{zero-cas}
e_1 f_2+h_1 h_2-2 f_1 e_2=0.
\end{gather}

\subsection[The 3 non-Abelian subcases of $\{\mathfrak{g}_2,\mathfrak{g}_2\}\subset\mathfrak{g}_1$]{The 3 non-Abelian subcases of $\boldsymbol{\{\mathfrak{g}_2,\mathfrak{g}_2\}\subset\mathfrak{g}_1}$}\label{g2g2=g1}

The relations (\ref{e2h2=ae1}) impose conditions on the functions $A$, $B$, $C$, giving $3$ subcases:
\begin{enumerate}\itemsep=0pt
 \item[1)] $A(r_0)=\sqrt{\frac{a}{2(r_0^2-1)}}$, $B(r_0)=0$, $C(r_0)=r_0A(r_0)$,
 \item[2)] $A(r_0)=0$, $B(r_0)=\sqrt{\frac{-a}{2}}$, $C(r_0)=0$, for $a<0$,
 \item[3)] $B(r_0)=0$, with $A(r_0)$ and $C(r_0)$ satisfying the equation
 \begin{gather}\label{eqAC}
 A'(r_0)-r_0 C'(r_0)=\frac{2A^2(r_0)-2 C^2(r_0)+a}{2r_0 A(r_0)-2 C(r_0)},\qquad\mbox{with}\qquad C(r_0)\neq r_0A(r_0).
 \end{gather}
\end{enumerate}

\subsubsection{Case 1:} Here we have the explicit solution (given here for $a=2$):
\begin{gather*}
 e_2 = \frac{q_1p_1+q_3p_3}{\sqrt{q_3^2-q_1^2}}, \qquad h_2 = \frac{2q_2(q_1p_1+q_3p_3)-2\big(q_3^2-q_1^2\big)p_2}{\sqrt{q_3^2-q_1^2}}, \\
 f_2 = \frac{2\big(q_1^2+q_2^2-q_3^2\big) (q_1p_1+q_3p_3)-4 q_2\big(q_3^2-q_1^2\big)p_2}{\sqrt{q_3^2-q_1^2}}.
\end{gather*}
In this case
\begin{gather*}
2 e_2 f_2-h_2^2 = -4 \left(e_1f_1+\frac{1}{4} h_1^2\right),
\end{gather*}
so the Casimir (\ref{Cas12}) vanishes, corresponding to a quadratic constraint between the basis elements.

The most general invariant of this $6$-dimensional algebra is a restriction of (\ref{g1-inv}), given by
$I_1 = F(r_0,r_1)$, with the most general quadratic invariant being
\begin{gather*}
H=\psi(r_0)r_1^2 = \psi\left(\frac{q_3}{q_1}\right)(q_3p_1+q_1p_3)^2, \qquad\mbox{with arbitrary function}\ \psi.
\end{gather*}

\subsubsection{Case 2:}

This just leads to the trivial case $\mathfrak{g}_2=\mathfrak{g}_1$.

\subsubsection{Case 3:}

This is the most interesting case, depending on two arbitrary functions, subject to one differential constraint (\ref{eqAC}). The explicit form of the Casimir (\ref{Cas12}) is
\begin{gather}
H = 2 \big(2(q_1A-q_3 C)^2+a \big(q_1^2-q_3^2\big)\big) \big(p_1^2-p_2^2-p_3^2\big) \nonumber\\
\hphantom{H =}{} + 2\big(a+2\big(A^2-C^2\big)\big) (q_3p_1+q_1p_3)^2,\label{Hg221case3}
\end{gather}
which is a specific example of the general quadratic integral (of $\mathfrak{g}_1$), given in (\ref{g1-I2}).

\br[constant curvature]
When
\begin{gather*}
C(r_0)\neq r_0A(r_0) \qquad \text{and} \qquad 2(A(r_0)-r_0C(r_0))^2 \neq a \big(r_0^2-1\big),
\end{gather*} then the matrix of coefficients is invertible and defines a metric with constant curvature, satisfying
\begin{gather}\label{concurv}
R_{ij}=\frac{1}{n} R g_{ij},
\end{gather}
where, in our case $n=3$ and $R=-12a$.
\er

The six first degree (in momenta) Hamiltonian functions generate six Killing vectors (by the formula~(\ref{3dVFs})) of the metric corresponding to the Hamiltonian (\ref{Hg221case3}). The Poisson algebra is given by Table~\ref{Tab:6Symm_alg}.

\begin{table}[h]\centering
\caption{The $6$-dimensional symmetry algebra $\mathfrak{g}$, when $\{\mathfrak{g}_2,\mathfrak{g}_2\}\subset\mathfrak{g}_1$.}\label{Tab:6Symm_alg}\vspace{1mm}
\begin{tabular}{|c||c|c|c||c|c|c|}
\hline &$e_1$ &$h_1$ &$f_1$ &$e_2$ &$h_2$ &$f_2$ \\[.10cm]\hline\hline
$e_1$ &0 &$2e_1$ &$-h_1$ &0 &$-2e_2$ &$-2h_2$ \\ \hline
$h_1$ &$-2e_1$ &0 &$2f_1$ &$-2e_2$ &0 &$2f_2$ \\ \hline
$f_1$ &$h_1$ &$-2f_1$ &0 &$-h_2$ &$-f_2$ &0 \\ \hline\hline
$e_2$ &0 &$2e_2$ &$h_2$ &0 &$ae_1$ &$-ah_1$ \\ \hline
$h_2$ &$2e_2$ &0 &$f_2$ &$-a e_1$ &0 &$-2a f_1$ \\ \hline
$f_2$ &$2h_2$ &$-2f_2$ &0 &$a h_1$ &$2a f_1$ &0 \\ \hline
\end{tabular}
\end{table}

\subsubsection*{The Lie algebra automorphisms $\boldsymbol{\iota_1}$ and $\boldsymbol{\iota_2}$}

The automorphisms of $\mathfrak{g}_1$, given by (\ref{i12}) also act on this extended algebra:
%\begin{subequations}\label{i12-g12}
\begin{gather*}
\iota_1\colon \ (e_1,h_1,f_1,e_2,h_2,f_2) \mapsto \left(f_1,-h_1,e_1,-\frac{1}{2} f_2,-h_2,-2 e_2\right),\\ % \label{i1-g12}\\
\iota_2\colon \ (e_1,h_1,f_1,e_2,h_2,f_2) \mapsto \left(-e_1,h_1,-f_1,e_2,-h_2,f_2\right), %\label{i2-g12}
\end{gather*}
with the Casimir function (\ref{Hg221case3}) being invariant (it being a function of the invariants $r_0$, $r_1$ and~$r_2$).

\subsection[The case when $\mathfrak{g}_2$ is Abelian]{The case when $\boldsymbol{\mathfrak{g}_2}$ is Abelian}\label{g2g2=0}

When $\mathfrak{g}_2$ is an Abelian algebra, we have
\begin{gather*}
 \{e_2,h_2\}=0,\qquad \{e_2,f_2\}=0,\qquad \{h_2,f_2\}=0,
\end{gather*}
so we have the Poisson algebra of Table \ref{Tab:6Symm_alg}, but with $a=0$, giving a $3\times 3$ block of zeros.

In this case the Killing form of the $6$-dimensional algebra $\mathfrak{g}$ is degenerate, but the Casimir can be obtained by taking the limit of (\ref{Cas12}) as $a\rightarrow 0$, giving
\begin{gather}\label{Cas0}
 H = 2 e_2 f_2 - h_2^2.
\end{gather}
As with Case 3, above, we have $B(r_0)=0$ and the functions $A(r_0)$ and $C(r_0)$ satisfy the differential constraint
\begin{gather}\label{eqAC0}
 A'(r_0)-r_0 C'(r_0)=\frac{A^2(r_0)-C^2(r_0)}{r_0 A(r_0)-C(r_0)},
\end{gather}
which is just (\ref{eqAC}) with $a=0$.

The explicit form of the Casimir (\ref{Cas0}) is given by
\begin{gather} \label{Hflat}
H = 4 (q_1A-q_3 C)^2 \big(p_1^2-p_2^2-p_3^2\big) + 4\big(A^2-C^2\big) (q_3p_1+q_1p_3)^2,
\end{gather}
which is just (\ref{Hg221case3}), with $a=0$, and non-degenerate when
\begin{gather}\label{detCas0}
(A- r_0 C) (C - r_0 A) \neq 0,
\end{gather}
in which case it corresponds to a {\it flat} metric when the functions $A$ and $C$ satisfy (\ref{eqAC0}).

\subsection{The solutions of (\ref{eqAC}) and (\ref{eqAC0})}\label{eqACAC0sols}

In Sections \ref{g2g2=g1} and \ref{g2g2=0}, we gave two classes of Poisson algebra $\mathfrak{g} = \mathfrak{g}_1 + \mathfrak{g}_2$, with Casimir functions~(\ref{Hg221case3}) and~(\ref{Hflat}), corresponding (when non-degenerate) to constant curvature and flat spaces, respectively. These depend on 2 functions $A(r_0)$ and $C(r_0)$, which must satisfy the differential relations~(\ref{eqAC}) or~(\ref{eqAC0}) respectively. In this section we consider the general solution of these equations and some particular cases of interest.

The general solution is constructed in two steps. First we reduce the problem to finding only one function $A(r_0)$, with $C(r_0)=1$ or $C(r_0)=0$. The second step reintroduces the second function.

\subsubsection{The solutions of (\ref{eqAC})}\label{eqACsols}

First, we note that $Z_i$ of (\ref{3dZi}) are only defined up to an overall multiple of a function of $r_0$, since this is an invariant of the algebra $\mathfrak{g}_1$. Therefore, to satisfy (\ref{3dZi}), we have two cases
\begin{enumerate}\itemsep=0pt
 \item[1)] $C(r_0)\neq 0$, in which case we may set $C(r_0)= 1$ and then determine the {\em one} function $A(r_0)$,
 \item[2)] $C(r_0)= 0$ $\Rightarrow$ $A(r_0)=\frac{1}{2}\sqrt{c_1r_0^2-2a}$.
\end{enumerate}
For the case $C(r_0)=1$, (\ref{eqAC}) takes the form
\begin{gather*}%\label{eqAC-C=0}
 A'(r_0) = \frac{2A^2(r_0)-2 + a}{2(r_0 A(r_0)-1)},\qquad\mbox{with}\qquad r_0A(r_0) \neq 1.
\end{gather*}
We then have a number of subcases.

When $A'(r_0)\neq 0$, we have the general solution
\begin{subequations}\label{C=1-A1}
\begin{gather}
A = \frac{(a-2)\big(2r_0 + c_1 \sqrt{2(a-2)} \sqrt{2-(a-2)^2 c_1^2+(a-2) r_0^2}\big)}{2 c_1^2 (a-2)^2 -4},\qquad\mbox{when} \ a\neq 2,\label{an=2}\\
A = \frac{1}{r_0\pm \sqrt{r_0^2-2 c_1}} = \frac{r_0\mp \sqrt{r_0^2-2 c_1}}{2c_1},\qquad\mbox{when}\ a = 2.\label{a=2}
\end{gather}
When $A'(r_0)= 0$, then
\begin{gather}\label{A=cons}
A(r_0)=\sqrt{\frac{2-a}{2}}.
\end{gather}
\end{subequations}
Clearly, when $(a-2)^2c_1^2-2=0$ the solution (\ref{an=2}) is singular. Replacing $c_1$ by $c_2=(a-2)^2c_1^2-2$, we find
\begin{subequations}\label{C=1-A2}
\begin{gather}
\big(2A^2+a-2\big) c_2 = (a-2)\big(4r_0A+(a-2)r_0^2-2\big) \nonumber\\
\qquad{}\Rightarrow\quad A = \frac{(2-a)r_0^2+2}{4r_0},\qquad \mbox{when}\ c_2=0.\label{singsol1}
\end{gather}
On the other hand, when $c_1=0$, we have the simple solution
\begin{gather}\label{singsol2}
A = \frac{(2-a) r_0}{2}.
\end{gather}
The first form of (\ref{a=2}) allows us to set $c_1=0$ (with the ``$+$'' sign) to obtain the special solution
\begin{gather}\label{singsol3}
A = \frac{1}{2r_0}.
\end{gather}
\end{subequations}

\subsubsection{The solutions of (\ref{eqAC0})}\label{eqAC0sols}

The solution of (\ref{eqAC0}) is just a reduction of those of (\ref{eqAC}), but with $a=0$, giving
\begin{subequations}\label{C=1-a=0}
\begin{gather}
A = \frac{r_0 + c_1 \sqrt{r_0^2+c_1^2-1}}{1- c_1^2}, \label{a=0gen}\\
A= \frac{r_0^2+1}{2r_0}, \label{a=0-singsol1} \\
A = r_0, \label{a=0-singsol2} \\
A = 1, \label{a=0-A=1}
\end{gather}
\end{subequations}
which are respectively reductions of (\ref{an=2}), (\ref{singsol1}), (\ref{singsol2}) and (\ref{A=cons}).

\subsubsection{Reinstating the second function}

We can now reinstate the second function by writing
\begin{gather}\label{add-sigma}
Z_1 = \sigma(r_0) (A(r_0) p_1+p_3),
\end{gather}
where $A(r_0)$ is one of the solutions (\ref{C=1-A1}) or (\ref{C=1-A2}). The conditions (\ref{e2h2=ae1}) then imply
\begin{gather}\label{sigma-eq}
\frac{2 \sigma \sigma'}{\sigma^2-1} = \frac{a}{(r_0-A)(r_0 A-1)},
\end{gather}
which can be directly integrated for a given solution $A(r_0)$.

We see from (\ref{sigma-eq}) that when $a=0$, we generically have $\sigma'=0$, so can just multiply the solutions (\ref{C=1-a=0}) by an arbitrary constant. There is a singular solution of (\ref{sigma-eq}): $ A=r_0$ and $\sigma$ {\em arbitrary}.

\subsection{The Casimirs (\ref{Hg221case3}) and (\ref{Hflat}) for some specific solutions}

The general formulae for the Casimirs (\ref{Hg221case3}) and (\ref{Hflat}) depend upon the specific functions $A$ and $C$. For any solution given in Section \ref{eqACAC0sols}, we can calculate the specific form of $H_0$ (the corresponding kinetic energy). Each one corresponds to a constant curvature or flat manifold, so will not all be independent. In fact, all constant curvature metrics with the same dimension, signature and scalar curvature $R$ are isometrically related (see \cite[p.~84]{97-8}). Since, in our case, we have $R=-12a$, any two cases with the same value of $a$ should be isometric, even though the transformation may be difficult to find.

\subsubsection{The Hamiltonian for case (\ref{A=cons})}

For this choice, (\ref{Hg221case3}) takes the form
\begin{subequations}\label{alg-A=cons}
\begin{gather}
H_0 = 2\big(\sqrt{2} q_1 - \sqrt{2-a} q_3\big)^2 \big(p_1^2-p_2^2-p_3^2\big) \label{H-A=cons} \\
\hphantom{H_0 =}{} = 4 q_1^2 \big(p_1^2-p_2^2-p_3^2\big), \qquad\mbox{when}\quad a=2. \label{H-a=2-A=cons}
\end{gather}
This restriction of $a=2$ corresponds to $A=0$, and gives the $6$-dimensional isometry algebra
\begin{gather}
 e_1 = p_2, \qquad h_1 = -2(q_1p_1+q_2p_2+q_3p_3), \nonumber\\
 f_1 = -2q_1q_2p_1+\big(q_3^2-q_1^2-q_2^2\big)p_2-2q_2q_3p_3 ,\nonumber\\
 e_2 = p_3, \qquad h_2 = 2(q_2p_3-q_3p_2), \nonumber\\
 f_2 = -4q_3(q_1p_1+q_2p_2)-2\big(q_1^2-q_2^2+q_3^2\big)p_3 , \label{6D-diag-alg}
\end{gather}
\end{subequations}
which satisfies the relations of Table~\ref{Tab:6Symm_alg} for $a=2$. This will be embedded into the $10$-dimensional algebra (\ref{diag-alg}) in Section~\ref{diagonal} and will be one of our main examples in the context of super-integrability in Section~\ref{Sect:super}.

\subsubsection{The Hamiltonian for case (\ref{singsol1})}

For this choice, (\ref{Hg221case3}) takes the form
\begin{gather*}
H_0 = \frac{\big(2 q_1^2+(a-2) q_3^2\big)^2}{4 q_1^2q_3^2} \big(\big(q_1^2+q_3^2\big)p_1^2-q_1^2p_2^2+2q_1q_3p_1p_3\big), %\label{H-singsol1}
\end{gather*}
which simplifies with the reduction $a=2$ and also reduces to the flat case, with $a=0$.

\subsubsection{The Hamiltonian for case (\ref{singsol2})}

For this choice, (\ref{Hg221case3}) takes the form
\begin{gather*}
H_0 = \big(2 q_1^2+(a-2) q_3^2\big) \left(a \big(p_1^2-p_2^2-p_3^2\big)+\left(\frac{a-2}{q_1^2}\right) \big(q_3p_1+q_1p_3\big)^2\right)\\ %\label{H-singsol2} \\
\hphantom{H_0}{} = 4 q_1^2 \big(p_1^2-p_2^2-p_3^2\big), \qquad\mbox{when}\quad a=2. %\label{H-a=2-singsol2}
\end{gather*}
This restriction to $a=2$ is identical to~(\ref{H-a=2-A=cons}), so corresponds to the same algebra~(\ref{6D-diag-alg}).

\subsubsection{The Hamiltonian for case (\ref{a=0-A=1})}

The flat case (\ref{a=0-A=1}) is just the case (\ref{A=cons}), with $a=0$, so $A=1$. However, we saw that when $a=0$, equation~(\ref{sigma-eq}) has a constant solution, so we make the choice $\sigma=\frac{1}{2}$, in which case, the Casimir~(\ref{Hflat}) takes the form
\begin{subequations}\label{alg-A=cons-a=0}
\begin{gather}
H_0 = (q_1 - q_3)^2 \big(p_1^2-p_2^2-p_3^2\big). \label{H-A=cons-a=0}
\end{gather}
The $6$-dimensional isometry algebra now takes the form
\begin{gather}
 e_1 = p_2, \qquad h_1 = -2(q_1p_1+q_2p_2+q_3p_3), \nonumber\\
 f_1 = -2q_1q_2p_1+\big(q_3^2-q_1^2-q_2^2\big)p_2-2q_2q_3p_3 ,\nonumber\\
e_2 = \frac{1}{2}(p_1+p_3), \qquad h_2 = q_2(p_1+p_3)+(q_1-q_3)p_2, \label{6D-diag-alg-a20}\\
f_2 = \big(q_2^2+(q_1-q_3)^2\big)p_1+2q_2(q_1-q_3)p_2+\big(q_2^2-(q_1-q_3)^2\big)p_3 , \nonumber
\end{gather}
\end{subequations}
which satisfies the relations of Table \ref{Tab:6Symm_alg} for $a=0$. This will be embedded into the $10$-dimensional algebra~(\ref{diag-alg-a20}) in Section~\ref{diagonala20} and will be one of our main examples in the context of super-integrability in Section~\ref{Sect:super}.

{\bf Flat coordinates.}
Since $e_2$, $h_2$, $f_2$ are in involution, we can consider them as new {\em momenta}, $P_1=e_2$, $P_2=h_2$, $P_3=f_2$, and find new coordinates $Q_i$, which are canonically conjugate. This is just Lie's theorem on complete integrability in the Poisson case. The equations $\{Q_i,P_j\}=\delta_{ij}$ give us a system of equations for~$Q_i$, which in the current case are easy to solve:
\begin{gather}\label{flat-coord}
Q_1 = \frac{q_1^2-q_2^2-q_3^2}{q_1-q_3},\qquad Q_2 = \frac{q_2}{q_1-q_3},\qquad Q_3 = \frac{-1}{2(q_1-q_3)}.
\end{gather}
With generating function $S=\frac{(q_1^2-q_2^2-q_3^2)P_1+q_2P_2-\frac{1}{2}P_3}{q_1-q_3}$, we then have
\begin{gather*}
e_1=-2(Q_2P_1+Q_3P_2),\qquad h_1=2(Q_3P_3-Q_1P_1),\qquad f_1=-Q_1P_2-Q_2P_3,\\ e_2=P_1,\qquad h_2=P_2,\qquad f_2=P_3,
\end{gather*}
leading to
\begin{gather}\label{H0-flat-coord}
H_0 = 2P_1P_3-P_2^2.
\end{gather}
The form of this is dictated by the form of the Casimir (\ref{Cas0}). It can, of course, be diagonalised to $H_0=2P_1^2-P_2^2-2P_3^2$ by using $Q_1\pm Q_3$.

\section{Extending to the conformal algebra}\label{Sect:conformal}

In Section \ref{Sect:PoisAlg} we built $6$-dimensional Poisson algebras which included $\mathfrak{g}_1$ as a subalgebra. The quadratic Casimir function was interpreted as a Hamiltonian function (the kinetic energy), with the algebra $\mathfrak{g} = \mathfrak{g}_1 + \mathfrak{g}_2$ being its {\em symmetry algebra}. When the matrix of coefficients was non-degenerate, this defined a metric, and the symmetry algebra corresponded to its {\em Killing vectors}. In this section we further extend the algebra $\mathfrak{g}$ to include conformal symmetries, which, in the metric case, correspond to {\em conformal Killing vectors}. In fact, we will first construct an extension with the appropriate Poisson bracket relations and then prove directly that these are conformal symmetries of the above Hamiltonian.

\subsection{Conformal algebras}

In 2 dimensions, as is well known, the conformal group is {\em infinite}. For
$n\ge 3$ this group is {\em finite} and has {\em maximal} dimension
$\frac{1}{2} (n+1)(n+2)$, which is achieved for {\em conformally flat} spaces
(which includes {\em flat} and {\em constant curvature} spaces). We are particularly interested in the case $n=3$, so will be looking for a $10$-dimensional algebra.

In flat spaces, the infinitesimal generators consist of $n$ {\em translations},
$\frac{1}{2} n (n-1)$ {\em rotations}, 1 {\em scaling} and $n$ {\em
inversions}, totalling $\frac{1}{2} (n+1)(n+2)$. This algebra is isomorphic to $\mathfrak{so}(n+1,1)$ (see \cite[p.~143]{84-4}).

For this discussion, we distinguish between ``true symmetries'', which we label $X_s$, and ``conformal symmetries'', which we label $X_c$. The ``true symmetries'' form a subalgebra of the conformal symmetry algebra. Here we discuss the general structure of the conformal algebra.

Suppose $X_s$ is a symmetry and $X_{c1}$, $X_{c2}$ are conformal symmetries of $H$, satisfying
\begin{gather*}
\{X_s,H\}=0,\qquad \{X_{ci},H\} = w_i H,
\end{gather*}
where $w_i$ are functions of the coordinates $q_1$, $q_2$, $q_3$. Then the Jacobi identity implies:
\begin{gather*}
\{\{X_s,X_{ci}\},H\} = - \{w_i,X_s\} H,\qquad \{\{X_{c1},X_{c2}\},H\} = (\{w_1,X_{c2}\} - \{w_2,X_{c1}\}) H.
\end{gather*}
The symmetry $X_s$ is, of course, just a special conformal symmetry, with $w=0$. Whilst it may be that $\{w_i,X_s\}=0$ for some particular choices of $X_s$ or $X_{ci}$ and that $(\{w_1,X_{c2}\} - \{w_2,X_{c1}\})$ may or may not be zero, these relations show that conformal symmetries form an invariant space under the action of the ``true'' symmetries and that the set of conformal symmetries (including the ``true'' symmetries) form a Lie algebra. In particular, the conformal symmetries must form an invariant space under the action of~$\mathfrak{g}_1$.

\subsection{Building the additional elements}

We start with the $6$-dimensional Lie algebra $\mathfrak{g} = \mathfrak{g}_1 + \mathfrak{g}_2$, where $\mathfrak{g}_2$ is either Case 3 of Section \ref{g2g2=g1} or the Abelian case of Section \ref{g2g2=0}. The respective Casimir functions $H$ correspond to a space with non-zero, constant curvature and a space of zero curvature.

We first {\em algebraically} extend $\mathfrak{g}$ by adding a further $4$ basis elements, so that, as a vector space, we have
\begin{gather*}
\hat{\mathfrak{g}} = \mathfrak{g}_1 + \mathfrak{g}_2 + \mathfrak{g}_3 + \mathfrak{g}_4,
\end{gather*}
where $\mathfrak{g}_3$ is another $3$-dimensional invariant space in the form of either Case 3 of Section \ref{g2g2=g1} or the Abelian case of Section \ref{g2g2=0}, and $\mathfrak{g}_4$ is a $1$-dimensional representation of the form (\ref{1drep}). We already know the bracket relations
\begin{gather*}
\{\mathfrak{g}_1,\mathfrak{g}_1\},\qquad \{\mathfrak{g}_2,\mathfrak{g}_2\},\qquad \{\mathfrak{g}_3,\mathfrak{g}_3\},\qquad \{\mathfrak{g}_1,\mathfrak{g}_2\},\qquad \{\mathfrak{g}_1,\mathfrak{g}_3\}\qquad \mbox{and}\qquad \{\mathfrak{g}_1,\mathfrak{g}_4\},
\end{gather*}
but need to derive
\begin{gather*}
\{\mathfrak{g}_2,\mathfrak{g}_3\},\qquad \{\mathfrak{g}_2,\mathfrak{g}_4\},\qquad \{\mathfrak{g}_3,\mathfrak{g}_4\}.
\end{gather*}
In fact, once we have determined the first of these, the remaining pair follow by the Jacobi identity.

We introduce the following notation for the basis elements of $\mathbf{\hat g}$:
\begin{gather*}
\mathfrak{g}_k = \{e_k,h_k,f_k\},\qquad \mbox{for}\quad k=1,2,3, \qquad\mbox{and}\qquad \mathfrak{g}_4 = \{h_4\},
\end{gather*}
with $\mathfrak{g}_1 + \mathfrak{g}_2$ satisfying the relations given by Table~\ref{Tab:6Symm_alg} (with $a=a_2$, possibly zero) and $\mathfrak{g}_1 + \mathfrak{g}_3$ satisfying the relations given by Table~\ref{Tab:6Symm_alg} (with $a=a_3$, possibly zero). We also have that~$h_4$ commutes with~$\mathfrak{g}_1$.

For $\mathbf{\hat g}$ to be a Lie algebra, we must have
\begin{gather*}
\{\mathfrak{g}_2,\mathfrak{g}_3\} \subset \mathfrak{g}_1 + \mathfrak{g}_2 + \mathfrak{g}_3 + \mathfrak{g}_4.
\end{gather*}
Noting that $\{\{e_2,h_3\},h_1\}= 2 \{e_2,h_3\}$, we have
\begin{gather*}
\{e_2,h_3\} = \alpha e_1 + \beta e_2 + \gamma e_3,\qquad\mbox{for arbitrary constants}\ \alpha,\ \beta,\ \gamma.
\end{gather*}
We can repeat the argument of Proposition~\ref{prop:g2g2} to show that, without loss of generality, we may choose $\beta = \gamma = 0$.
Defining
\begin{gather*}
\hat e_2 = e_2 + \mu e_1 \quad \Rightarrow \quad \hat h_2 = h_2 - \mu h_1 \qquad\mbox{and}\qquad \hat f_2 = f_2 - 2\mu f_1, \\
\hat e_3 = e_3 + \nu e_1 \quad \Rightarrow \quad \hat h_3 = h_3 - \nu h_1 \qquad\mbox{and}\qquad \hat f_3 = f_3 - 2\nu f_1,
\end{gather*}
for arbitrary parameters $\mu$, $\nu$, then
\begin{gather*}
\{\hat e_2,\hat h_3\} = (\alpha-2 \mu\nu) e_1 + (\beta-2\nu) e_2 + (\gamma-2\mu) e_3.
\end{gather*}
Choosing $\mu=\frac{1}{2} \gamma$, $\nu=\frac{1}{2} \beta$, we have
\begin{gather*}
\{\hat e_2,\hat h_3\} = \hat \alpha e_1,\qquad\mbox{where}\qquad \hat \alpha = \alpha -\frac{1}{2} \beta \gamma.
\end{gather*}
Dropping ``hats'', we have shown that $\{e_2,h_3\} = a_4 e_1$, for some parameter~$a_4$. The next proposition extends this to the whole of $\{\mathfrak{g}_2,\mathfrak{g}_3\}$, as shown in Table~\ref{Tab:g2_g3}.

\bp\label{prop:g2g3}
Bases can be chosen for $\mathfrak{g}_2$ and $\mathfrak{g}_3$, satisfying $\{\mathfrak{g}_2,\mathfrak{g}_3\}\subset\mathfrak{g}_1+\mathfrak{g}_4$. Specifically, there exist parameters $a_4$, $\gamma$, such that the relations shown in Table~{\rm \ref{Tab:g2_g3}} are satisfied.
\begin{table}[h]\centering
\caption{The relations for $\{\mathfrak{g}_2,\mathfrak{g}_3\}$.}\label{Tab:g2_g3}\vspace{1mm}
\begin{tabular}{|c||c|c|c|}
\hline &$e_3$ &$h_3$ &$f_3$ \\[.10cm]\hline\hline
$e_2$ &0 & $a_4 e_1$ & $\gamma h_4-a_4 h_1$ \\ \hline
$h_2$ & $-a_4 e_1$ &$-\gamma h_4$ & $-2a_4 f_1$ \\ \hline
$f_2$ & $\gamma h_4+a_4 h_1$ & $2a_4 f_1$ & 0 \\ \hline
\end{tabular}
\end{table}
\ep

\begin{proof}
First, we note that since $\{\{e_2,e_3\},h_1\}=4 \{e_2,e_3\}$, we have $\{e_2,e_3\}=0$. Similarly, we find $\{f_2,f_3\}=0$. We then have
\begin{gather*}
\{\{e_2,e_3\},f_1\}=0 \quad \Rightarrow \quad \{h_2,e_3\}=-\{e_2,h_3\} = -a_4 e_1, \\
\{\{f_2,f_3\},e_1\}=0 \quad \Rightarrow \quad \{h_2,f_3\}+\{f_2,h_3\} = 0.
\end{gather*}
Further action of $f_1$ leads to
\begin{gather*}
\{h_2,h_3\}+\{e_2,f_3\}=-a_4 h_1 \qquad\mbox{and}\qquad \{f_2,e_3\} + \{h_2,h_3\}= a_4 h_1.
\end{gather*}
Since $\{\{h_2,h_3\},h_1\}=0$, we have $\{h_2,h_3\}=-\gamma h_4+\delta h_1$, so, bracketing this with $f_1$ gives
\begin{gather*}
\{h_2,f_3\}+\{f_2,h_3\} = 2 \delta f_1 \quad\Rightarrow\quad \delta = 0.
\end{gather*}
Piecing these results together, we obtain Table~\ref{Tab:g2_g3}.
\end{proof}

Now that we have $\{\mathfrak{g}_2,\mathfrak{g}_3\}$, we calculate $\{\mathfrak{g}_2,\mathfrak{g}_4\}$ and $\{\mathfrak{g}_3,\mathfrak{g}_4\}$ by using the Jacobi identity. We require $\gamma\neq 0$ if $\mathfrak{g}_4$ is to enter our calculations, so, without loss of generality, we may take $\gamma=1$, but leave $a_4$ arbitrary.

Since $\{h_2,h_3\}=- h_4$, we have
\begin{gather*}
\{e_2,h_4\} = - \{e_2,\{h_2,h_3\}\} = \{h_2,\{h_3,e_2\}\}+\{h_3,\{e_2,h_2\}\}=2 (a_2e_3-a_4 e_2),
\end{gather*}
using the relations we already have. Similarly, we can derive the remaining brackets to complete Table \ref{Tab:10conf_alg}. The lower part of the table is, of course, determined by skew symmetry.
\begin{table}[h]\centering
\caption{The 10-dimensional conformal algebra when $\{\mathfrak{g}_i,\mathfrak{g}_i\}\subset\mathfrak{g}_1$.}\label{Tab:10conf_alg}\vspace{1mm}
{\footnotesize
\begin{tabular}{|c||c|c|c||c|c|c||c|c|c||c|}
\hline &$e_1$ &$h_1$ &$f_1$ &$e_2$ &$h_2$ &$f_2$ &$e_3$ &$h_3$ &$f_3$ &$h_4$\\[.10cm]\hline\hline
$e_1$ &0 &$2e_1$ &$-h_1$ &0 &$-2e_2$ &$-2h_2$ &0 &$-2e_3$ &$-2h_3$ &0\\ \hline
$h_1$ & &0 &$2f_1$ &$-2e_2$ &0 &$2f_2$ &$-2e_3$ &0 &$2f_3$ &0\\ \hline
$f_1$ & & &0 &$-h_2$ &$-f_2$ &0 &$-h_3$ &$-f_3$ &0 &0\\ \hline\hline
$e_2$ & & & &0 &$a_2e_1$ &$-a_2h_1$ &0 & $a_4e_1$ &$h_4-a_4h_1$ &$2(a_2e_3-a_4e_2)$\\ \hline
$h_2$ & & & & &0 &$-2a_2f_1$ & $-a_4e_1$ &$-h_4$ & $-2a_4f_1$ &$2 (a_2h_3-a_4h_2)$\\ \hline
$f_2$ & & & & & &0 &$h_4+a_4h_1$ & $2a_4f_1$ &0 &$2(a_2f_3-a_4f_2)$\\ \hline\hline
$e_3$ & & & & & & &0 &$a_3e_1$ &$-a_3h_1$ &$2(a_4e_3-a_3e_2)$\\ \hline
$h_3$ & & & & & & & &0 &$-2a_3f_1$ &$2(a_4h_3-a_3h_2)$\\ \hline
$f_3$ & & & & & & & & &0 &$2(a_4f_3-a_3f_2)$\\ \hline\hline
$h_4$ & & & & & & & & & &0\\ \hline
\end{tabular}
}\end{table}

The cases for which $\{\mathfrak{g}_2,\mathfrak{g}_2\}=\textbf{0}$ and/or $\{\mathfrak{g}_3,\mathfrak{g}_3\}=\textbf{0}$ are obtained by setting $a_2=0$ and/or $a_3=0$.

\subsubsection*{The Lie algebra automorphisms $\boldsymbol{\iota_1}$ and $\boldsymbol{\iota_2}$}

The automorphisms of $\mathfrak{g}_1$, given by (\ref{i12}) also act on this $10$-dimensional algebra:
\begin{gather*}
\begin{array}{|c||c|c|c|c|c|c|c|c|c|c|c|c|c|}\hline
& e_1 & h_1 & f_1 & e_2 & h_2 & f_2 & e_3 & h_3 & f_3 & h_4 \\ \hline\hline
\iota_1\colon & f_1 & -h_1 & e_1 & -\frac{1}{2} f_2 & -h_2 & -2 e_2 & -\frac{1}{2}f_3 & -h_3 & -2 e_3 & h_4\\[1.5mm] \hline
\iota_{23}\colon & -e_1 & h_1 & -f_1 & e_2 & -h_2 & f_2 & e_3 & -h_3 & f_3 & h_4\\ \hline
\end{array}
\end{gather*}
Note that the four spaces $\mathfrak{g}_1$, $\mathfrak{g}_2$, $\mathfrak{g}_3$ and $\mathfrak{g}_4$ are each invariant.

\subsubsection{The equations for the coefficients}

Table \ref{Tab:10conf_alg} was obtained from Table \ref{Tab:g2_g3} by requiring {\em algebraic consistency} as an abstract Poisson algebra. However, these Poisson relations impose additional {\it differential} relations on the functions used to define the basis elements. We will solve the resulting equations for~$(A_3,C_3)$ in terms of~$(A_2,C_2)$, which will be arbitrary solutions of equations (\ref{eqAC}) or (\ref{eqAC0}).

From Case 3 of Section~\ref{g2g2=g1} or the Abelian case of Section \ref{g2g2=0}, we have
\begin{gather*}
 e_i = A_ip_1+C_ip_3, \\
 h_i = 2A_iq_2p_1+2(A_iq_1-C_iq_3)p_2+2C_iq_2p_3,\\ %\label{Zij} \\
 f_i = 2\big(A_i\big(q_1^2+q_2^2+q_3^2\big)-2C_iq_1q_3\big)p_1+4q_2(A_iq_1-C_iq_3)p_2 \\
\hphantom{f_i =}{} +2\big(C_i\big(q_2^2-q_1^2-q_3^2\big)+2A_iq_1q_3\big)p_3,
\end{gather*}
where $i=2, 3$ and $A_i(r_0)$, $C_i(r_0)$ satisfy either (\ref{eqAC}) (with parameter $a\rightarrow a_i$) or (\ref{eqAC0}), as well as
\begin{gather*}%\label{Z41}
 h_4 = C_4(r_0) (q_3p_1+q_1p_3),
\end{gather*}
as the basis of $\mathfrak{g}_4$.

We must solve the two equations
\begin{subequations}\label{e2h2h3}
\begin{gather}
\{e_2,h_3\} = a_4 e_1, \label{e2h3}\\
 \{h_2,h_3\} = -h_4, \label{h2h3}
\end{gather}
\end{subequations}
each of which has $3$ components (the coefficients of $p_i$).

Equations (\ref{eqAC}) (or (\ref{eqAC0})), together with the $p_3$ component of (\ref{e2h3}), can be used to eliminate the derivatives $A_i'(r_0)$ and $C_i'(r_0)$ (for $i=2,3$), and then the $p_3$ component of (\ref{h2h3}) gives the formula
\begin{gather*}%\label{C4}
C_4 = 4(A_2C_3-A_3C_2).
\end{gather*}
It is then possible to solve the remaining parts of (\ref{e2h2h3}) for $A_3(r_0)$ and $C_3(r_0)$, but the solution depends upon whether or not $a_2a_3=0$.

\subsection[The case $\{\mathfrak{g}_2,\mathfrak{g}_2\}\subset\mathfrak{g}_1$ and $\{\mathfrak{g}_3,\mathfrak{g}_3\}\subset\mathfrak{g}_1$]{The case $\boldsymbol{\{\mathfrak{g}_2,\mathfrak{g}_2\}\subset\mathfrak{g}_1}$ and $\boldsymbol{\{\mathfrak{g}_3,\mathfrak{g}_3\}\subset\mathfrak{g}_1}$}\label{gigi=g1}

When $a_2a_3\neq 0$, we obtain
\begin{gather}
A_3(r_0) = \frac{\sqrt{a_2a_3-a_4^2} (2r_0A_2C_2-2A_2^2-a_2)}{a_2\sqrt{2\big(a_2 \big(r_0^2-1\big)-2 (r_0C_2-A_2)^2\big)}} + \frac{a_4 A_2}{a_2}, \nonumber\\
 C_3(r_0) = \frac{\sqrt{a_2a_3-a_4^2} (2r_0C_2^2-2A_2C_2-a_2r_0)}{a_2\sqrt{2\big(a_2 \big(r_0^2-1\big)-2 (r_0C_2-A_2)^2\big)}} + \frac{a_4 C_2}{a_2}, \label{Case1Sol1} \\
 C_4(r_0) = 4(A_2C_3-A_3C_2)=\frac{4\sqrt{a_2a_3-a_4^2} (C_2-r_0A_2)}{\sqrt{2\big(a_2 \big(r_0^2-1\big)-2 (r_0C_2-A_2)^2\big)}}, \nonumber
\end{gather}
where $A_2(r_0)$, $C_2(r_0)$ are arbitrary solutions of equation (\ref{eqAC}) with $a=a_2$.

\subsubsection{Casimir functions and conformal factors}\label{cascon}

Table \ref{Tab:10conf_alg} can be rearranged by re-ordering the $4$ subspaces of $\mathbf{\hat g}$. We can take
\begin{itemize}\itemsep=0pt
 \item $\mathfrak{g}_1+\mathfrak{g}_2$ as Killing vectors of a Casimir $H_{12}$, with conformal Killing vectors in the space $\mathfrak{g}_3+\mathfrak{g}_4$.
 \item $\mathfrak{g}_1+\mathfrak{g}_3$ as Killing vectors of a Casimir $H_{13}$, with conformal Killing vectors in the space $\mathfrak{g}_2+\mathfrak{g}_4$.
 \item $\mathfrak{g}_1+\mathfrak{g}_4$ as Killing vectors of a Casimir $H_{14}$, with conformal Killing vectors in the space $\mathfrak{g}_2+\mathfrak{g}_3$.
\end{itemize}
Since they all have the same $10$-dimensional conformal algebra, they are conformally equivalent to one another.

{\it The Hamiltonian $H_{12}$} will denote the Casimir corresponding to the sub-algebra $\mathfrak{g}_1+\mathfrak{g}_2$, and is given by (\ref{Hg221case3}), but with $(a,A,C)=(a_2,A_2,C_2)$. This corresponds to a metric of {\it constant curvature}, with $R=-12a_2$. The $6$-dimensional algebra $\mathfrak{g}_1+\mathfrak{g}_2$ is just the symmetry algebra and $\mathfrak{g}_3+\mathfrak{g}_4$ correspond to conformal symmetries, satisfying
\begin{gather*}
\{e_3,H_{12}\}=w_{31} H_{12},\qquad \{h_3,H_{12}\}=w_{32} H_{12},\\
 \{f_3,H_{12}\}=w_{33} H_{12},\qquad \{h_4,H_{12}\}=w_{34} H_{12}.
\end{gather*}
We need to calculate $w_{31}$ directly, obtaining
\begin{gather*}
w_{31}=\frac{a_2a_3-a_4^2}{a_2(q_1A_3-q_3C_3)-a_4(q_1A_2-q_3C_2)},
\end{gather*}
but the remaining (infinitesimal) conformal factors can be derived by using the Poisson bracket relations of Table~\ref{Tab:10conf_alg}:
\begin{gather*}
w_{32}= \{w_{31},f_1\}=2 q_2 w_{31},\qquad w_{33}= \{w_{32},f_1\} = 2\big(q_2^2+q_3^2-q_1^2\big) w_{31},
\end{gather*}
and
\begin{gather*}
\{h_4,H_{12}\}=\{\{h_3,h_2\},H_{12}\}=\{\{h_3,H_{12}\},h_2\}=\{w_{32},h_2\}H_{12} \\
\qquad{} \Rightarrow\quad w_{34}=4 (q_1A_2-q_3C_2) w_{31}.
\end{gather*}
\br\label{w3i-rep}
The $3$ functions $w_{3i}$ form a representation space for our algebra $\mathfrak{g}_1$. Under the action of the Poisson bracket, we have
\begin{gather*}
 f_1\colon \ (w_{31},w_{32},w_{33})\mapsto (w_{32},w_{33},0),\qquad h_1\colon \ (w_{31},w_{32},w_{33})\mapsto (2w_{31},0,-2w_{33}), \\
 e_1\colon \ (w_{31},w_{32},w_{33})\mapsto (0,2w_{31},2w_{32}).
\end{gather*}
The function $w_{34}$ is invariant with respect to $\mathfrak{g}_1$.
\er

{\it The Hamiltonian $H_{13}$} corresponds to the sub-algebra $\mathfrak{g}_1+\mathfrak{g}_3$ and is again of the form (\ref{Hg221case3}), but now with $(a,A,C)=(a_3,A_3,C_3)$, so corresponds to a metric of {\it constant curvature}, with $R=-12a_3$. The $6$-dimensional algebra $\mathfrak{g}_1+\mathfrak{g}_3$ is now the symmetry algebra and $\mathfrak{g}_2+\mathfrak{g}_4$ correspond to conformal symmetries.

Since $H_{13}$ has the same conformal algebra as $H_{12}$, the corresponding metrics must be conformally related. To see this (on the level of the inverse metric) we use formulae (\ref{Case1Sol1}) to replace $A_3, C_3$ in $H_{13}$ to obtain
\begin{gather*}
H_{13}=\phi_{13} H_{12},\quad\mbox{where}\quad \phi_{13} = \phi_{13}^0 + \frac{2 a_4 \sqrt{2\big(a_2a_3-a_4^2\big)}}{a_2^2} \phi_{13}^1+ \frac{a_4^2}{a_2^2} \phi_{13}^2,
\end{gather*}
where
\begin{gather*}
 \phi_{13}^0 = \frac{2 a_3 (q_1A_2-q_3 C_2)^2}{a_2 \big(a_2\big(q_3^2-q_1^2\big)-2(q_1A_2-q_3 C_2)^2\big)},\\
 \phi_{13}^1 = \frac{q_3 C_2-q_1A_2}{\sqrt{a_2 \big(a_2\big(q_3^2-q_1^2\big)-2(q_1A_2-q_3 C_2)^2\big)}},\\
 \phi_{13}^2 = \frac{a_2\big(q_3^2-q_1^2\big)-4(q_1A_2-q_3 C_2)^2}{a_2\big(q_3^2-q_1^2\big)-2(q_1A_2-q_3 C_2)^2}.
\end{gather*}
Since $\{e_2,H_{12}\}=0$, we have
\begin{gather*}
\{e_2,H_{13}\}=\{e_2,\log(\phi_{13})\}H_{13},
\end{gather*}
giving
\begin{gather*}
w_{21}=\{e_2,\log(\phi_{13})\}\\
\hphantom{w_{21}}{} = \frac{a_2\sqrt{2 \big(a_2a_3-a_4^2\big)}}{\sqrt{2 \big(a_2a_3-a_4^2\big)} (q_1A_2-q_3C_2)-a_4\sqrt{a_2\big(q_3^2-q_1^2\big)-2(q_1A_2-q_3 C_2)^2}}.
\end{gather*}
Again, with the notation $\{h_2,H_{13}\}=w_{22} H_{13}$, $\{f_2,H_{13}\}=w_{23} H_{13}$, we use the action of $f_1$ to find
\begin{gather*}
w_{22}= 2 q_2 w_{21},\qquad w_{23}= 2\big(q_2^2+q_3^2-q_1^2\big) w_{21}.
\end{gather*}
We can then use $h_4 = \{h_3,h_2\}$ to obtain
\begin{gather*}
w_{24} = -\frac{2}{a_2} \Big(2 a_4 (q_1A_2-q_3C_2)+\sqrt{2 \big(a_2a_3-a_4^2\big)}\sqrt{a_2\big(q_3^2-q_1^2\big)-2(q_1A_2-q_3 C_2)^2}\Big) w_{21},
\end{gather*}
where $\{h_4,H_{13}\} = w_{24}H_{13}$.

{\it The Hamiltonian $H_{14}$} corresponds to the sub-algebra $\mathfrak{g}_1+\mathfrak{g}_4$. Since $\mathfrak{g}_4$ contains the single element $h_4$, defined by $Z_{41}$ of (\ref{1drep}), with $C_4$ given by (\ref{Case1Sol1}), and since $\{\mathfrak{g}_1,\mathfrak{g}_4\}=\{\mathfrak{g}_4,\mathfrak{g}_4\}={\bf 0}$, this is an algebraically trivial extension, since it is just a {\it direct sum}. However, the Casimir,
\begin{gather*}
H_{14} = e_1f_1+\frac{1}{4} h_1^2+ \alpha h_4^2 = \big(r_0^2+\alpha C_4^2(r_0)\big)r_1^2+\big(r_0^2-1\big) r_2 \\
\hphantom{H_{14} =}{} = \big(q_1^2+\alpha q_3^2 C_4^2\big)p_1^2+2q_1q_3\big(1+\alpha C_4^2\big)p_1p_3+\big(q_3^2-q_1^2\big)p_2^2+\big(q_3^2+\alpha q_1^2C_4^2\big)p_3^2,
\end{gather*}
defines a non-degenerate upper-index metric whenever $\alpha \neq 0$, which is conformally equivalent to $H_{12}$ when $\alpha=\frac{1}{4(a_2a_3-a_4^2)}$, satisfying
\begin{gather*}
H_{14}=\phi_{14} H_{12},\qquad\mbox{where}\qquad \phi_{14} = \frac{q_3^2-q_1^2}{2a_2\big(q_3^2-q_1^2\big)-4(q_1A_2-q_3 C_2)^2}.
\end{gather*}
The metric, corresponding to $H_{14}$, has constant {\it scalar} curvature $R=-2$, but is not actually a~constant curvature metric, since it does not satisfy (\ref{concurv}) and, indeed, only has a $4$-dimensional symmetry algebra.

The elements of $\mathfrak{g}_2+\mathfrak{g}_3$ correspond to conformal symmetries of $H_{14}$. Defining $z_{ki}$ by
\begin{gather*}
\{e_k,H_{14}\}=z_{k1} H_{14},\qquad \{h_k,H_{14}\}=z_{k2} H_{14},\qquad \{f_k,H_{14}\}=z_{k3} H_{14},\qquad\mbox{for}\quad k=2,3,
\end{gather*}
we again have $z_{21}=\{e_2,\log(\phi_{14})\}$ and use the action of $f_1$ to find
\begin{gather*}
z_{21}=\frac{2(q_1A_2-q_3C_2)}{q_3^2-q_1^2},\qquad z_{22}= 2 q_2 z_{21},\qquad z_{23}= 2\big(q_2^2+q_3^2-q_1^2\big) z_{21}.
\end{gather*}
Noting that
\begin{gather*}
\{e_3,H_{14}\} = \{e_3,\log(\phi_{14})\} H_{14}+\phi_{14}\{e_3,H_{12}\} \quad\Rightarrow\quad z_{31}= \{e_3,\log(\phi_{14})\} + w_{31},
\end{gather*}
we find that $z_{3i}$ are given by the same formulae as $z_{2i}$, but with $(A_2,C_2)$ replaced by $(A_3,C_3)$.

\br
The function $\phi_{14}$ satisfies
\begin{gather*}
\phi_{14} = \frac{1}{8\big(a_2a_3-a_4^2\big)} \left(w_{31}w_{33}-\frac{1}{2} w_{32}^2\right),
\end{gather*}
which is an invariant of the representation mentioned in Remark \ref{w3i-rep}.
\er

\subsubsection*{The Lie algebra automorphisms $\boldsymbol{\iota_1}$ and $\boldsymbol{\iota_2}$}

Under the action of $\iota_1$ of (\ref{i1}), we have
\begin{gather*}
\frac{q_3}{q_1} \mapsto \frac{q_3}{q_1}, \qquad q_1^2-q_2^2-q_3^2 \mapsto \frac{1}{q_1^2-q_2^2-q_3^2},\qquad (H_{12},H_{13},H_{14})\mapsto (H_{12},H_{13},H_{14}).
\end{gather*}
For $k=2,3$, the functions
\begin{gather*}
(w_{k1},w_{k2},w_{k3},w_{k4})\mapsto \left(-\frac{1}{2} w_{k3},-w_{k2},-2 w_{k1}, w_{k4}\right) \qquad\mbox{and}\qquad (\phi_{13},\phi_{14})\mapsto (\phi_{13},\phi_{14}),
\end{gather*}
and similarly for $z_{ki}$. The action of $\iota_2$ is even simpler.

These automorphisms will be very important in later sections, when we discuss super-integrable systems associated with some of our Casimir functions.

\subsubsection{Reduction to the diagonal case}\label{diagonal}

Consider the Hamiltonian $H_{12}$, which is of the form (\ref{Hg221case3}), but with $(a,A,C)=(a_2,A_2,C_2)$. The only off-diagonal term is the coefficient of $p_1p_3$, which vanishes when $2A_2^2-2C_2^2+a_2=0$, which then implies that the right-hand side of~(\ref{eqAC}) also vanishes, so we have
\begin{gather*}
A_2'-r_0C_2'=0,\qquad A_2A_2'-C_2C_2'=0 \quad\Rightarrow\quad (r_0 A_2-C_2) C_2'=0.
\end{gather*}
Since we require that $r_0 A_2-C_2\neq 0$, we have
\begin{gather*}
C_2=c_1 \quad \mbox{(a constant)}\quad\Rightarrow\quad A_2 = \sqrt{\frac{2c_1^2-a_2}{2}},
\end{gather*}
so
\begin{gather*}
H_{12} = 2\Big(\sqrt{2} c_1 q_1-\sqrt{2 c_1^2-a_2} q_3\Big)^2 \big(p_1^2-p_2^2-p_3^2\big) = 2a_2q_1^2\big(p_1^2-p_2^2-p_3^2\big),
\end{gather*}
when $c_1$ is chosen so that $2c_1^2=a_2$. This is just the case of equations (\ref{alg-A=cons}). For the choice $a_2=-a_3=2, a_4=0$, the conformal algebra has the explicit form:
\begin{gather}
e_1 = p_2, \qquad h_1 = -2(q_1p_1+q_2p_2+q_3p_3), \nonumber\\
 f_1 = -2q_1q_2p_1+\big(q_3^2-q_1^2-q_2^2\big)p_2-2q_2q_3p_3 ,\nonumber\\
e_2 = p_3, \qquad h_2 = 2(q_2p_3-q_3p_2), \qquad f_2 = -4q_3(q_1p_1+q_2p_2)-2\big(q_1^2-q_2^2+q_3^2\big)p_3 , \label{diag-alg}\\
 e_3 = p_1, \qquad h_3 = 2(q_1p_2+q_2p_1), \qquad f_3 = 2\big(q_1^2+q_2^2+q_3^2\big)p_1+4q_1(q_2p_2+q_3p_3),\nonumber\\
 h_4=-4(q_3p_1+q_1p_3),\nonumber
\end{gather}
which is a $10$-dimensional extension of the algebra (\ref{6D-diag-alg}).
In this case we have
\begin{gather}
H_{12}= 4 q_1^2\big(p_1^2-p_2^2-p_3^2\big),\qquad H_{13}= 4 q_3^2\big(p_1^2-p_2^2-p_3^2\big),\nonumber\\ H_{14}= \big(q_1^2-q_3^2\big)\big(p_1^2-p_2^2-p_3^2\big).\label{H1234}
\end{gather}

\br[further automorphism]
As can be seen, the Casimir $H_{12}$ is invariant under the interchange $2\leftrightarrow 3$, which induces the following involution $\iota_{23}$ of the $10$-dimensional algebra
\begin{gather}\label{i23}
(e_1,h_1,f_1,e_2,h_2,f_2,e_3,h_3,f_3,h_4) \mapsto \left(e_2,h_1,\frac{1}{2} f_2,e_1,-h_2,2 f_1,e_3,-\frac{1}{2} h_4,f_3,-2h_3\right).
\end{gather}
This is no longer an automorphism of $\mathfrak{g}_1$, so its representation spaces are not individually preserved, but it is an automorphism of the symmetry algebra $\mathfrak{g}=\mathfrak{g}_1+\mathfrak{g}_2$ and of the conformal elements $\mathfrak{g}_3+\mathfrak{g}_4$.
\er

\subsubsection[$H_{12}$ of (\ref{H1234}) as a reduction from flat space in 4-dimensions]{$\boldsymbol{H_{12}}$ of (\ref{H1234}) as a reduction from flat space in 4-dimensions}\label{diagonal-4d}

Starting with $\mathfrak{g}_1$ of (\ref{diag-alg}), we can build a $3$-dimensional, highest weight representation in the space of functions of $q_1$, $q_2$, $q_3$. We obtain
\begin{gather*}
y_1=\frac{1}{q_1},\qquad y_2 = \frac{2q_2}{q_1},\qquad y_3 = \frac{2\big(q_2^2+q_3^2-q_1^2\big)}{q_1},
\end{gather*}
satisfying $\{(y_1,y_2,y_3),e_1\}=(0,2y_1,2y_2)$, $\{(y_1,y_2,y_3),h_1\}=(2y_1,0,-2y_3)$, $\{(y_1,y_2,y_3),f_1\}=(y_2,y_3,0)$. When acting on these with $\mathfrak{g}_2$, we need to add the function $y_4=\frac{2q_3}{q_1}$, which Poisson commutes with the whole of $\mathfrak{g}_1$ (it is just $2r_0$ (see (\ref{g1-inv}))). The action of $\mathfrak{g}_2$ is given by
\begin{gather*}
\{(y_1,y_2,y_3,y_4),e_2\}=(0,0,2y_4,2y_1),\qquad \{(y_1,y_2,y_3,y_4),h_2\}=(0,-2y_4,0,2y_2),\\ \{(y_1,y_2,y_3,y_4),f_2\}=(2y_4,0,0,2y_3).
\end{gather*}
These clearly define a linear action of $\mathfrak{g}_1+\mathfrak{g}_2$ on $\{y_i\}_{i=1}^4$, given by
\begin{gather*}
T_xf=\{f,x\},\qquad\mbox{satisfying}\qquad [T_x,T_y]f = - T_{\{x,y\}}f.
\end{gather*}
The four variables $y_i$ satisfy the quadratic constraint $2y_1y_3-y_2^2-y_4^2=-4$, which defines a~quadratic form with matrix
\begin{gather*}
S= \left(
 \begin{matrix}
 0 & 0 & 1 & 0 \\
 0 & -1 & 0 & 0 \\
 1 & 0 & 0 & 0 \\
 0 & 0 & 0 & -1 \\
 \end{matrix}
 \right),
\end{gather*}
and the matrices $T_x$ for $x\in \mathfrak{g}_1+\mathfrak{g}_2$ are ``infinitesimally orthogonal'' with respect to this ``metric'', satisfying $T_x S + S T_x^t=0$. This means that our symmetry algebra $\mathfrak{g}_1+\mathfrak{g}_2$ is just $\mathfrak{so}(1,3)$. If we use the matrix $S$ to define the corresponding Lorentzian metric, we find
\begin{gather*}
{\rm d}s^2 = 2 {\rm d}y_1{\rm d}y_3-{\rm d}y_2^2-{\rm d}y_4^2 = \frac{4}{q_1^2} \big({\rm d}q_1^2-{\rm d}q_2^2-{\rm d}q_3^2\big),
\end{gather*}
corresponding to the Hamiltonian $H_{12}$ of (\ref{H1234}).

\subsection[The case $\{\mathfrak{g}_2,\mathfrak{g}_2\}={\bf 0}$ and $\{\mathfrak{g}_3,\mathfrak{g}_3\}\subset\mathfrak{g}_1$]{The case $\boldsymbol{\{\mathfrak{g}_2,\mathfrak{g}_2\}={\bf 0}}$ and $\boldsymbol{\{\mathfrak{g}_3,\mathfrak{g}_3\}\subset\mathfrak{g}_1}$}\label{g2g2=0g3}

Here we must solve equations (\ref{e2h2h3}) with $a_2=0$, so $A_2(r_0)$, $C_2(r_0)$ satisfy equation (\ref{eqAC0}), while $A_3(r_0)$, $C_3(r_0)$ satisfy equation~(\ref{eqAC}), with $a=a_3$, which can be either zero or non-zero. The calculation soon gives the choice of
\begin{itemize}\itemsep=0pt
 \item $a_4=0$, leading to $A_2 = r_0 C_2$, which means that the determinant condition (\ref{detCas0}) is not satisfied, so the Casimir function (\ref{Cas0}) cannot be associated with a flat metric.
 \item $a_4\neq 0$, which leads to a non-degenerate flat metric, but has a {\em singular limit} as $a_4\rightarrow 0$. This is the only case we consider here.
\end{itemize}
When $a_4\neq 0$ we find
\begin{gather*}
 A_3 = \frac{a_3 A_2}{2a_4}+\frac{a_4\big(2q_1q_3C_2-\big(q_1^2+q_3^2\big)A_2\big)}{4 (q_1A_2-q_3C_2)^2},\qquad C_3 = \frac{a_3 C_2}{2a_4}+\frac{a_4\big(\big(q_1^2+q_3^2\big)C_2-2q_1q_3A_2\big)}{4 (q_1A_2-q_3C_2)^2}, \\
% \label{A3C3a2=0} \\
 C_4 = 4(A_2C_3-A_3C_2)= \frac{2 a_4 (q_1C_2-q_3 A_2)}{q_1A_2-q_3 C_2}, \nonumber
\end{gather*}
where $A_2(r_0)$, $C_2(r_0)$ are arbitrary solutions of equation (\ref{eqAC0}).

\subsubsection{Casimir functions and conformal factors}

We now have Table~\ref{Tab:10conf_alg}, with $a_2=0$ and again consider various $6$-dimensional subalgebras and their respective Casimir functions. The automorphisms~(\ref{i12}) are still valid in this case.

{\it The Hamiltonian $H_{12}$} will again denote the Casimir corresponding to the sub-algebra $\mathfrak{g}_1+\mathfrak{g}_2$, and is given by~(\ref{Hflat}), but with $(A,C)=(A_2,C_2)$. This corresponds to a {\it flat} metric. The $6$-dimensional algebra $\mathfrak{g}_1+\mathfrak{g}_2$ is just the symmetry algebra and $\mathfrak{g}_3+\mathfrak{g}_4$ correspond to conformal symmetries, satisfying
\begin{gather*}
\{e_3,H_{12}\}=w_{31} H_{12},\qquad \{h_3,H_{12}\}=w_{32} H_{12},\\ \{f_3,H_{12}\}=w_{33} H_{12},\qquad \{h_4,H_{12}\}=w_{34} H_{12}.
\end{gather*}
The coefficients are calculated in the same way to give
\begin{gather*}
w_{31}=\frac{a_4}{q_1A_2-q_3C_2},\qquad w_{32}= 2 q_2 w_{31},\qquad w_{33}= 2\big(q_2^2+q_3^2-q_1^2\big) w_{31}, \qquad w_{34}=4 a_4.
\end{gather*}

{\it The Hamiltonian $H_{13}$} corresponds to the sub-algebra $\mathfrak{g}_1+\mathfrak{g}_3$ and is of the form (\ref{Hg221case3}), with $(a,A,C)=(a_3,A_3,C_3)$, so corresponds to a metric of {\it constant curvature}, with $R=-12a_3$. The $6$-dimensional algebra $\mathfrak{g}_1+\mathfrak{g}_3$ is now the symmetry algebra and $\mathfrak{g}_2+\mathfrak{g}_4$ correspond to conformal symmetries.

Again $H_{13}$ is conformally related to $H_{12}$, with
\begin{gather*}
H_{13}=\phi_{13} H_{12},\qquad\mbox{where}\qquad \phi_{13} = \frac{\big(2 a_3 (q_1A_2-q_3C_2)^2-a_4^2\big(q_3^2-q_1^2\big)\big)^2}{16 a_4^2 (q_1A_2-q_3C_2)^4}
\end{gather*}
Defining $w_{2k}$ by
\begin{gather*}
\{e_2,H_{13}\}=w_{21} H_{13},\qquad \{h_2,H_{13}\}=w_{22} H_{13},\\ \{f_2,H_{13}\}=w_{23} H_{13},\qquad \{h_4,H_{13}\}=w_{24} H_{13},
\end{gather*}
we have
\begin{gather*}
 w_{21}=\frac{4a_4^2(q_1A_2-q_3C_2)}{a_4^2\big(q_3^2-q_1^2\big)-2 a_3 (q_1A_2-q_3C_2)^2},\qquad w_{22}= 2 q_2 w_{21},\\
 w_{23}= 2\big(q_2^2+q_3^2-q_1^2\big) w_{21}, \qquad w_{24}= - \frac{a_4^2\big(q_3^2-q_1^2\big)+2 a_3 (q_1A_2-q_3C_2)^2}{a_4(q_1A_2-q_3C_2)} w_{21}.
\end{gather*}

{\it The Hamiltonian $H_{14}$} corresponds to the sub-algebra $\mathfrak{g}_1+\mathfrak{g}_4$, and is given by
\begin{gather*}
H_{14} = e_1f_1+\frac{1}{4} h_1^2+ \alpha h_4^2 = \big(r_0^2+\alpha C_4^2(r_0)\big)r_1^2+\big(r_0^2-1\big) r_2 \\
\hphantom{H_{14}}{} = \big(q_1^2+\alpha q_3^2 C_4^2\big)p_1^2+2q_1q_3\big(1+\alpha C_4^2\big)p_1p_3+\big(q_3^2-q_1^2\big)p_2^2+\big(q_3^2+\alpha q_1^2C_4^2\big)p_3^2,
\end{gather*}
which is non-degenerate whenever $\alpha \neq 0$, and is conformally equivalent to $H_{12}$ when $\alpha=\frac{-1}{4a_4^2}$, satisfying
\begin{gather*}
H_{14}=\phi_{14} H_{12},\qquad\mbox{where}\qquad \phi_{14} = - \frac{q_3^2-q_1^2}{4(q_1A_2-q_3 C_2)^2}.
\end{gather*}
As before, the metric, corresponding to $H_{14}$, has constant {\it scalar} curvature $R=-2$, but is not actually a constant curvature metric, since it does not satisfy~(\ref{concurv}) and, indeed, only has a~$4$-dimensional symmetry algebra.

The elements of $\mathfrak{g}_2+\mathfrak{g}_3$ correspond to conformal symmetries of~$H_{14}$. Defining $z_{ki}$ by
\begin{gather*}
\{e_k,H_{14}\}=z_{k1} H_{14},\qquad \{h_k,H_{14}\}=z_{k2} H_{14},\qquad \{f_k,H_{14}\}=z_{k3} H_{14},\qquad\mbox{for}\quad k=2,3,
\end{gather*}
we use the action of $f_1$ to find
\begin{gather*}
z_{21}=\frac{2(q_1A_2-q_3C_2)}{q_3^2-q_1^2},\qquad z_{22}= 2 q_2 z_{21},\qquad z_{23}= 2\big(q_2^2+q_3^2-q_1^2\big) z_{21},
\end{gather*}
with $z_{3i}$ being given by the same formulae as $z_{2i}$, but with $(A_2,C_2)$ replaced by $(A_3,C_3)$.

\subsubsection{Reduction to the diagonal case}\label{diagonala20}

The diagonalisation of the Hamiltonian $H_{12}$ is simpler in the flat case. The only off-diagonal term is the coefficient of $p_1p_3$, which now vanishes when $A_2^2-C_2^2=0$, so $C_2 = \pm A_2$, corresponding to
\begin{gather*}
H_{12} = 4 (q_1\mp q_3)^2 A_2^2 \big(p_1^2-p_2^2-p_3^2\big).
\end{gather*}
For simplicity, we choose $C_2=A_2=\frac{1}{2}$, after which we find
\begin{gather}
 H_{12} = (q_1 - q_3)^2 \big(p_1^2-p_2^2-p_3^2\big), \qquad
 H_{13} = \left(a_4(q_1+q_3) +\frac{a_3}{2a_4} (q_1 - q_3)\right)^2 \big(p_1^2-p_2^2-p_3^2\big), \nonumber\\
H_{14} = \big(q_1^2 - q_3^2\big) \big(p_1^2-p_2^2-p_3^2\big). \label{Hija2=0}
\end{gather}
The conformal algebra now has the explicit form:
\begin{gather}
e_1 = p_2, \quad h_1 = -2(q_1p_1+q_2p_2+q_3p_3), \qquad f_1 = -2q_1q_2p_1+\big(q_3^2-q_1^2-q_2^2\big)p_2-2q_2q_3p_3 ,\nonumber\\
 e_2 = \frac{1}{2}(p_1+p_3), \qquad h_2 = q_2(p_1+p_3)+(q_1-q_3)p_2, \nonumber\\
f_2 = \big(q_2^2+(q_1-q_3)^2\big)p_1+2q_2(q_1-q_3)p_2+\big(q_2^2-(q_1-q_3)^2\big)p_3 , \label{diag-alg-a20}\\
 e_3 = -\frac{1}{2}a_4(p_1-p_3)+\frac{a_3}{2a_4} e_2, \qquad h_3 = -a_4(q_2(p_1-p_3)+(q_1+q_3)p_2)+\frac{a_3}{2a_4} h_2, \nonumber\\
 f_3 = -a_4\big(\big((q_1+q_3)^2+q_2^2\big)p_1+2q_2(q_1+q_3)p_2+\big((q_1+q_3)^2-q_2^2\big)p_3\big)+\frac{a_3}{2a_4} f_2,\nonumber\\
 h_4= 2a_4(q_3p_1+q_1p_3).\nonumber
\end{gather}
This is an extension of the algebra given in (\ref{alg-A=cons-a=0}).

\br
As can be seen, this algebra is no longer invariant under the involution $\iota_{23}$, given by~(\ref{i23}).
\er

\section{Adding potentials: separability}\label{Sect:separable}

In this section we consider Hamiltonian systems of the form
\begin{gather*}
H = H_0 + h({\boldsymbol{q}}),
\end{gather*}
with the kinetic energy $H_0$ being one of our diagonal cases of $H_{12}$, given by (\ref{H1234}) or~(\ref{Hija2=0}), $I_2$ (of~(\ref{g1-I2})) or~$H_{14}$, given by~(\ref{H1234}).

For complete integrability (in the Liouville sense) we need two functions $F_1$ and $F_2$, such that $H$, $F_1$, $F_2$ are in involution:
\begin{gather}\label{involution}
\{H,F_1\} = \{H,F_2\} = \{F_1,F_2\} = 0.
\end{gather}
We restrict attention to functions $F_i$, whose dependence on momenta is at most quadratic. Such functions will be the sum of two homogeneous parts, $F_i=F^{(2)}_i + F^{(0)}_i$, and
\begin{gather*}
\{H,F_i\}=0 \quad\Rightarrow\quad \big\{H_0,F^{(2)}_i\big\}=0 \qquad\mbox{and}\qquad \big\{H_0,F^{(0)}_i\big\}+\big\{h,F^{(2)}_i\big\} = 0.
\end{gather*}
The first of these means that the coefficients of~$p_i p_j$ in~$F^{(2)}_i$ define a second order Killing tensor of the metric corresponding to~$H_0$. When this metric is constant curvature, {\it all} Killing tensors are built as tensor products of Killing vectors (see~\cite{74-7}). In the Poisson representation, this just means that~$F^{(2)}_i$ is some quadratic form of the elements of $\mathfrak{g}=\mathfrak{g}_1+\mathfrak{g}_2$ (of Section~\ref{diagonal} or Section~\ref{diagonala20}). Since, in each case, this algebra is of rank $2$, any $K\in \mathfrak{g}$ will commute with exactly one other element~$\bar K$. Since we require $\big\{F^{(2)}_1,F^{(2)}_2\big\}=0$, we must choose these quadratic parts to be independent quadratic form of some pair~$K$, $\bar K$. For simplicity, we will choose our pairs to be one of $e_1$, $e_2$, or $h_1$, $h_2$ or $f_1$, $f_2$. In the conformally flat case of~$H_0=I_2$, we have a smaller symmetry algebra, but since $I_2$ is no longer the Casimir function, we can use~$H_0$,~${\cal C}_1$ and a choice of~$K^2$ to generate our integrals.

The choice of quadratic integrals means that our systems will be separable. The calculation of separable potentials is standard and it is well known that in the standard orthogonal coordinate systems, with separable kinetic energies, we can add potentials which depend upon a number of arbitrary functions of a single variable~\cite{76-8}. If a complete (possessing $n$ parameters) solution of the Hamilton--Jacobi equation is found, then, by Jacobi's theorem, these parameters, when written in terms of the canonical variables, are quadratic (in momenta) first integrals of~$H$. The problem has also been posed in the ``opposite'' direction: given a pair of Poisson commuting, homogeneously quadratic integrals (in two degrees of freedom) what sort of potentials can be added, whilst maintaining commutativity? This is a classical problem (see Whittaker \cite[Chapter~12, Section~152]{88-4}) and leads to the Bertrand--Darboux equation for the potential~\cite{88-8,08-7}. This approach will be used in this section. The calculations are very similar, so the details will be omitted (with a few more included in the first case).

\subsection{The constant curvature case of (\ref{H1234})}\label{concurv-sep}

Here we consider the Hamiltonian
\begin{gather*}%\label{H0+h}
H = H_0 + h({\boldsymbol{q}}), \qquad\mbox{where}\qquad H_0=q_1^2\big(p_1^2-p_2^2-p_3^2\big),
\end{gather*}
with the kinetic energy $H_0=\frac{1}{4}H_{12}$ of (\ref{H1234}) and the specific conformal algebra (\ref{diag-alg}).

\subsubsection[The commuting pair $h_1$, $h_2$]{The commuting pair $\boldsymbol{h_1}$, $\boldsymbol{h_2}$}

Consider the case of
\begin{gather*}
F_1 = \frac{1}{4} h_1^2+g_1({\boldsymbol{q}})=(q_1p_1+q_2p_2+q_3p_3)^2+g_1({\boldsymbol{q}}),\nonumber\\
 F_2 = \frac{1}{4} h_2^2+g_2({\boldsymbol{q}})= (q_2p_3-q_3p_2)^2+g_2({\boldsymbol{q}}).%\label{h^2}
\end{gather*}
Each of the equations (\ref{involution}) is {\it linear} in momenta, so give us $9$ equations in all. This is an overdetermined system for the $3$ functions $h$, $g_1$, $g_2$, which can be solved explicitly in terms of~$3$ functions, each of a single variable. We find
\begin{gather*}
\pa_1 g_2=0,\qquad (q_2\pa_2+q_3\pa_3)g_2=0 \quad\Rightarrow\quad g_2 = \varphi_2\left(\frac{q_3}{q_2}\right),
\end{gather*}
after which
\begin{gather*}
q_2 (q_2\pa_2+q_3\pa_3)g_1=(q_1\pa_1+q_2\pa_2+q_3\pa_3)g_2=0 \quad \Rightarrow \quad g_1=\Phi_1\big(q_1,q_2^2+q_3^2\big), \\
q_3 (q_2\pa_2+q_3\pa_3)h= q_1^2 \pa_2 g_2 \quad \Rightarrow \quad h=\psi\big(q_1,q_2^2+q_3^2\big)-\frac{q_1^2 \varphi_2\big(\frac{q_3}{q_2}\big)}{q_2^2+q_3^2}.
\end{gather*}
There are $2$ more independent equations, leading to
\begin{gather*}
(q_2\pa_1+q_1\pa_2)\Phi_1=0 \quad \Rightarrow \quad \Phi_1\big(q_1,q_2^2+q_3^2\big) = \varphi_1\big(q_1^2-q_2^2-q_3^2\big), \\
(q_1\pa_1+q_2\pa_2+q_3\pa_3) h = q_1\pa_1 \Phi_1 \\
\qquad{} \Rightarrow \quad
 \psi\big(q_1,q_2^2+q_3^2\big) = \frac{q_1^2 \varphi_1\big(q_1^2-q_2^2-q_3^2\big)}{q_1^2-q_2^2-q_3^2} + \varphi_3\left(\frac{q_2^2+q_3^2}{q_1^2}\right) .
\end{gather*}
In summary, we have
\begin{gather*}
 h = \frac{q_1^2 \varphi_1\big(q_1^2-q_2^2-q_3^2\big)}{q_1^2-q_2^2-q_3^2} -\frac{q_1^2 \varphi_2\big(\frac{q_3}{q_2}\big)}{q_2^2+q_3^2}+ \varphi_3\left(\frac{q_2^2+q_3^2}{q_1^2}\right), \\
%\label{hhsols1} \\
 g_1 = \varphi_1\big(q_1^2-q_2^2-q_3^2\big), \qquad g_2 = \varphi_2\left(\frac{q_3}{q_2}\right).
\end{gather*}
This solution immediately gives us the separation variables
\begin{gather*}
u = q_1^2-q_2^2-q_3^2,\qquad v = \frac{q_3}{q_2},\qquad w = \frac{q_2^2+q_3^2}{q_1^2}\\
\qquad{}\Rightarrow\quad
 h= \frac{\varphi_1(u)}{1-w}-\frac{\varphi_2(v)}{w}+ \varphi_3(w),\qquad g_1 = \varphi_1(u), \qquad g_2 = \varphi_2(v).
\end{gather*}

\br[action of automorphism]
Under the action of the automorphism $\iota_1$ of (\ref{i1}), we have
\begin{gather*}
(u,v,w) \mapsto \left(\frac{1}{u},-v,w\right),
\end{gather*}
so this solution is invariant up to redefining some arbitrary functions.
\er

\subsubsection[The commuting pairs $e_1$, $e_2$ and $f_1$, $f_2$]{The commuting pairs $\boldsymbol{e_1}$, $\boldsymbol{e_2}$ and $\boldsymbol{f_1}$, $\boldsymbol{f_2}$}

These two cases are connected by the action of the automorphism $\iota_1$, of (\ref{i1}). The simplest case to calculate is with the pair $e_1$, $e_2$:
\begin{subequations}\label{e1e2}
\begin{gather}\label{e^2}
F_1 = e_1^2+g_1({\boldsymbol{q}})= p_2^2+g_1({\boldsymbol{q}}),\qquad F_2 = e_2^2+g_2({\boldsymbol{q}})= p_3^2+g_2({\boldsymbol{q}}).
\end{gather}
The simple form of $e_1$ and $e_2$ means that we are already in separation coordinates, leading to
\begin{gather}\label{eesols1}
h = -q_1^2(\varphi_1(q_2)+\varphi_2(q_3))+\varphi_3(q_1),\qquad g_1 = \varphi_1(q_2),\qquad g_2 = \varphi_2(q_3).
\end{gather}
\end{subequations}

The much more difficult case to calculate, involving $f_1$ and $f_2$, is simply obtained by using the automorphism $\iota_1$, which preserves $H_0$, whilst mapping $e_1^2\mapsto f_1^2$ and $e_2^2\mapsto \frac{1}{4} f_2^2$. This gives
%\begin{subequations}
\begin{gather*}
 F_1 = f_1^2+g_3({\boldsymbol{q}})= \big({-}2q_1q_2p_1+\big(q_3^2-q_1^2-q_2^2\big)p_2-2q_2q_3p_3\big)^2+g_3({\boldsymbol{q}}), \\
 %\label{f^2} \\
 F_2 = \frac{1}{4}f_2^2+g_4({\boldsymbol{q}})= \big(2q_3(q_1p_1+q_2p_2)+\big(q_1^2-q_2^2+q_3^2\big)p_3\big)^2+g_4({\boldsymbol{q}}),
\end{gather*}
where
\begin{gather*}
g_3 = \varphi_1\left(\frac{-q_2}{q_1^2-q_2^2-q_3^2}\right), \qquad g_4 = \varphi_2\left(\frac{q_3}{q_1^2-q_2^2-q_3^2}\right),\\
% \label{g34} \\
 h = \varphi_3\left(\frac{q_1}{q_1^2-q_2^2-q_3^2}\right) - \frac{q_1^2 \left(\varphi_1\left(\frac{-q_2}{q_1^2-q_2^2-q_3^2}\right)+\varphi_2\left(\frac{q_3}{q_1^2-q_2^2-q_3^2}\right)\right)}{\big(q_1^2-q_2^2-q_3^2\big)^2}.
\end{gather*}

\subsection{The flat case of (\ref{Hija2=0})}

Here we consider the Hamiltonian
\begin{gather}\label{H0flat+h}
H = H_0 + h({\boldsymbol{q}}), \qquad\mbox{where}\qquad H_0= (q_1-q_3)^2\big(p_1^2-p_2^2-p_3^2\big),
\end{gather}
with the kinetic energy $H_0=H_{12}$ of (\ref{Hija2=0}) and the specific conformal algebra (\ref{diag-alg-a20}).

\subsubsection[The commuting pair $h_1$, $h_2$]{The commuting pair $\boldsymbol{h_1}$, $\boldsymbol{h_2}$}\label{sec:flat-h1h2}

Consider the case of
\begin{gather*}%\label{h^2flat}
F_1 = \frac{1}{4} h_1^2+g_1({\boldsymbol{q}})=(q_1p_1+q_2p_2+q_3p_3)^2+g_1({\boldsymbol{q}}),\\ F_2=h_2^2+g_2({\bf{q}})= (q_2(p_1+p_3)+(q_1-q_3)p_2)^2+g_2(\bf{q}).
\end{gather*}
The relations $\{H,F_1\}=\{H,F_2\}=\{F_1,F_2\}=0$ lead to
\begin{gather*}
 h =\frac{ (q_1-q_3)^2\varphi_1\big(q_1^2-q_2^2-q_3^2\big)}{q_1^2-q_2^2-q_3^2}-\varphi_2\left(\frac{q_1-q_3}{q_2}\right) +\varphi_3\left(\frac{2q_1^2-2q_1q_3-q_2^2}{(q_1-q_3)^2}\right),\\
g_1 = \varphi_1\big(q_1^2-q_2^2-q_3^2\big),\qquad g_2 = \varphi_2\left(\frac{q_1-q_3}{q_2}\right).
\end{gather*}
Again, we have the separation variables
\begin{gather*}
u = q_1^2-q_2^2-q_3^2,\qquad v = \frac{q_1-q_3}{q_2},\qquad w = \frac{2q_1^2-2q_1q_3-q_2^2}{(q_1-q_3)^2} ,
\end{gather*}
with
\begin{gather*}
h= \frac{\varphi_1(u)}{w-1}-\varphi_2(v)+ \varphi_3(w),\qquad g_1 = \varphi_1(u), \qquad g_2 = \varphi_2(v),
\end{gather*}
and again we have
$(u,v,w) \mapsto \left(\frac{1}{u},-v,w\right)$, under the automorphism $\iota_1$, so the solution is invariant up to redefining some arbitrary functions.

\subsubsection[The commuting pairs $e_1$, $e_2$ and $f_1$, $f_2$]{The commuting pairs $\boldsymbol{e_1}$, $\boldsymbol{e_2}$ and $\boldsymbol{f_1}$, $\boldsymbol{f_2}$}\label{sec:flat-e1e2}

These two cases are again connected by the action of the automorphism $\iota_1$. The simplest case to calculate is with the pair $e_1$, $e_2$:
\begin{gather*}%\label{e^2flat}
F_1=e_1^2+g_1({\bf{q}})=p_2^2+g_1({\bf{q}}),\qquad F_2=4e_2^2+g_2({\bf{q}})=(p_1+p_3)^2+g_2(\bf{q}).
\end{gather*}
The relations $\{H,F_1\}=\{H,F_2\}=\{F_1,F_2\}=0$ lead to
\begin{gather*}
h = -(q_1-q_3)^2\varphi_1(q_2)+q_1(q_1-q_3)^2 \varphi_2'(q_1-q_3)+\varphi_3(q_1-q_3), \\
g_1 = \varphi_1(q_2),\qquad g_2 = \varphi_2(q_1-q_3).
\end{gather*}

Again, the case involving $f_1$ and $f_2$ is simply obtained by using the automorphism $\iota_1$, which preserves $H_0$, and still maps $e_1^2\mapsto f_1^2$ and $e_2^2\mapsto \frac{1}{4} f_2^2$. This gives
\begin{gather*}
 F_1=f_1^2+g_3({\bf{q}})=\big({-}2q_1q_2p_1+\big(q_3^2-q_1^2-q_2^2\big)p_2-2q_2q_3p_3\big)^2+g_3({\bf{q}}), \\
 %\label{f^2flat} \\
 F_2=f_2^2+g_4({\bf{q}})=\big(\big(q_2^2+(q_1-q_3)^2\big)p_1+2q_2(q_1-q_3)p_2+\big(q_2^2-(q_1-q_3)^2\big)p_3\big)^2+g_4(\bf{q}),
\end{gather*}
where
\begin{gather*}
 g_3 = \varphi_1\left(\frac{q_2}{q_1^2-q_2^2-q_3^2}\right), \qquad g_4 = \varphi_2\left(\frac{q_1-q_3}{q_1^2-q_2^2-q_3^2}\right), \\
 %\label{g34flat} \\
 h = -\frac{(q_1-q_3)^2\varphi_1\left(\frac{q_2}{q_1^2-q_2^2-q_3^2}\right)}{\big(q_1^2-q_2^2-q_3^2\big)^2}+\frac{q_1(q_1-q_3)^2 \varphi_2'\left(\frac{q_1-q_3}{q_1^2-q_2^2-q_3^2}\right)}{\big(q_1^2-q_2^2-q_3^2\big)^3}
 +\varphi_3\left(\frac{q_1-q_3}{q_1^2-q_2^2-q_3^2}\right).
\end{gather*}

\subsection{Conformally flat cases}\label{conflat-sep}

In this section we consider $H_0$ to be the diagonal case of $I_2$ (see (\ref{g1-I2}))
\begin{gather}\label{I2+h}
H = H_0 + h({\boldsymbol{q}}), \qquad\mbox{where}\qquad H_0= \varphi\left(\frac{q_3}{q_1}\right) q_1^2\big(p_1^2-p_2^2-p_3^2\big).
\end{gather}
We exclude the case $\varphi(r_0)=(c_1r_0+c_2)^2$, since this corresponds to the constant curvature case of Section~\ref{diagonal}. Generally, this $H_0$ has only~3 symmetries~($\mathfrak{g}_1$), but in the case $\varphi=c_1(r_0^2-1)$ it has a fourth symmetry and corresponds to the case $H_0=H_{14}$ of~(\ref{H1234}), with symmetry algebra $\mathfrak{g}_1+\mathfrak{g}_4$.

For generic $\varphi(r_0)$, the kinetic energy $H_0$ has the symmetry algebra~$\mathfrak{g}_1$. It is easy to check that $H_0$, ${\cal C}_1$ (of~(\ref{C1})) and $K^2$ (for any element $K$ of $\mathfrak{g}_1$) are functionally independent, so we use these to construct some associated involutive systems, with
\begin{gather}\label{gen-A}
H = H_0 + h,\qquad F_1 = {\cal C}_1 + g_1,\qquad F_2 = K^2 +g_2,
\end{gather}
where $h$, $g_1$, $g_2$ are arbitrary functions of $q_1$, $q_2$, $q_3$.

We just present the results. The calculations are straightforward.

\subsubsection*{The Case $\boldsymbol{K=e_1}$}

Involutivity of (\ref{gen-A}) leads to
\begin{gather}
h = \varphi_1(r_0)+\frac{q_1^2 \varphi(r_0)}{q_1^2-q_3^2} \big(\varphi_2\big(q_3^2-q_1^2\big)+\big(q_3^2-q_1^2\big)\varphi_3(q_2)\big),\nonumber\\
g_1 = \varphi_2\big(q_3^2-q_1^2\big)+\big(q_3^2-q_1^2\big)\varphi_3(q_2),\qquad g_2 = \varphi_3(q_2), \label{K=e1}
\end{gather}
which gives the separation variables $u = \frac{q_3}{q_1}$, $v = q_3^2-q_1^2$, $w = q_2$, in terms of which
\begin{gather*}
H=\big(u^2-1\big) \varphi(u) p_u^2+\varphi_1(u)-\frac{\varphi(u)}{u^2-1} F_1,\\
 F_1 = 4 v^2 p_v^2+\varphi_2(v) + v F_2, \qquad F_2 = p_w^2+\varphi_3(w).
\end{gather*}
\br[the involution $\iota_1$]
We can use the involution $\iota_1$ to transform this system to an equivalent one for which $K=f_1$.
\er

\subsubsection*{The case $\boldsymbol{K=h_1}$}

Involutivity of (\ref{gen-A}) leads to
\begin{gather}
 h = \varphi_1(r_0)+\frac{q_1^2 \varphi(r_0)}{q_1^2-q_3^2} \left(\varphi_2\left(\frac{q_3^2-q_1^2}{q_2^2}\right)+\frac{q_1^2-q_3^2}{4\big(q_1^2-q_2^2-q_3^2\big)} \varphi_3\big(q_1^2-q_2^2-q_3^2\big)\right),\nonumber\\
g_1 = \varphi_2\left(\frac{q_3^2-q_1^2}{q_2^2}\right)+\frac{q_1^2-q_3^2}{4\big(q_1^2-q_2^2-q_3^2\big)} \varphi_3\big(q_1^2-q_2^2-q_3^2\big),\qquad g_2 = \varphi_3\big(q_1^2-q_2^2-q_3^2\big),\label{K=h1}
\end{gather}
which gives the separation variables $u = \frac{q_3}{q_1}$, $v = \frac{q_3^2-q_1^2}{q_2^2}$, $w = q_1^2-q_2^2-q_3^2$, in terms of which
\begin{gather*}
H=(u^2-1) \varphi(u) p_u^2+\varphi_1(u)-\frac{\varphi(u)}{u^2-1} F_1,\nonumber\\
F_1 = 4 v^2(v+1) p_v^2+\varphi_2(v) + \frac{v}{4(v+1)} F_2, \qquad F_2 = 16 w^2p_w^2+\varphi_3(w). \nonumber
\end{gather*}
\br[the involution $\iota_1$]
This system is invariant under the action of the involution $\iota_1$, up to a relabelling of $\varphi_3$, since $q_1^2-q_2^2-q_3^2\mapsto \frac{1}{q_1^2-q_2^2-q_3^2}$.
\er

\subsubsection[The conformally flat case $H_{14}$ of (\ref{H1234})]{The conformally flat case $\boldsymbol{H_{14}}$ of (\ref{H1234})}\label{sec:H14}

The kinetic energy $H_0=H_{14}$ of~(\ref{H1234}) is a specific example of $H_0$ of (\ref{I2+h}), corresponding to $\varphi(r_0)=1-r_0^2$, and has the $4$-dimensional symmetry algebra $\mathfrak{g}_1+\mathfrak{g}_4$, with basis $e_1$, $h_1$, $f_1$, $h_4$, with~$h_4$ commuting with the whole of $\mathfrak{g}_1$. Consequently the cases~(\ref{K=e1}) and~(\ref{K=h1}) simply reduce to this choice of $\varphi(r_0)$. However, there are additional possibilities involving the element~$h_4$. Since $H_0 = {\cal C}_1-\frac{1}{16} h_4^2$, we cannot use~(\ref{gen-A}), with $K=h_4$. We can, however, use $H_0$, $h_4$ and any element~$K$ of~$\mathfrak{g}_1$.

\subsubsection*{The commuting pair $\boldsymbol{h_4}$, $\boldsymbol{e_1}$}\label{sec:H14-h4e1}

With the choice
\begin{gather*}
F_1 = \frac{1}{16} h_4^2+g_1({\boldsymbol{q}}), \qquad F_2 = e_1^2+g_2({\boldsymbol{q}}),
\end{gather*}
a simple calculation leads to
\begin{gather*}
 h = \varphi_3\big(q_3^2-q_1^2\big) + \big(q_3^2-q_1^2\big) \varphi_2(q_2) - \varphi_1\left(\frac{q_3}{q_1}\right),
\end{gather*}
depending upon $3$ arbitrary functions, with $g_1=\varphi_1\left(\frac{q_3}{q_1}\right)$ and $g_2=\varphi_2(q_2)$.

We can use $\iota_1$ to derive an equivalent system with $e_1$ replaced by $f_1$.

\subsubsection*{The commuting pair $\boldsymbol{h_4}$, $\boldsymbol{h_1}$}\label{sec:H14-h4h1}

With the choice
\begin{gather*}
F_1 = \frac{1}{16} h_4^2+g_1({\boldsymbol{q}}), \qquad F_2 = \frac{1}{4} h_1^2+g_2({\boldsymbol{q}}),
\end{gather*}
a simple calculation leads to
\begin{gather*}
 h = \varphi_3\left(\frac{q_3^2-q_1^2}{q_2^2}\right) + \frac{\big(q_3^2-q_1^2\big)\varphi_2\big(q_2^2+q_3^2-q_1^2\big)}{q_2^2+q_3^2-q_1^2} - \varphi_1\left(\frac{q_3}{q_1}\right),
\end{gather*}
depending upon $3$ arbitrary functions, with $g_1=\varphi_1\big(\frac{q_3}{q_1}\big)$ and $g_2=\varphi_2\big(q_2^2+q_3^2-q_1^2\big)$.

This system is invariant (up to a simple redefinition of~$\varphi_2$) under the action of~$\iota_1$.

\section{Adding potentials: super-integrability}\label{Sect:super}

In this section we consider the possibility of adding further integrals, $F_3$, $F_4$, to separable cases of Section~\ref{Sect:separable}, which can no longer be in involution with $H$, $F_1$, $F_2$, but should Poisson commute with $H$ itself: $\{H,F_3\}=\{H,F_4\}=0$. Having any additional integrals, the system is referred to as {\em super-integrable}. The functions should be chosen to be {\em functionally independent}, so the Jacobian matrix
\begin{gather*}
\frac{\pa(H,F_i)}{\pa {\boldsymbol{x}}},\qquad\mbox{where}\qquad {\boldsymbol{x}} = (q_1,\dots , p_3),
\end{gather*}
has {\it maximal rank}. Whilst the maximal rank for a set of functions in this space is $6$, the maximal rank for a set of {\it first integrals} is $5$, since in this case, the level surface
\begin{gather*}%\label{levelcurve}
{\cal S} = \{{\boldsymbol{x}}\colon H=c_0,\,F_i=c_i\}_{i=1}^4,
\end{gather*}
has dimension {\it one}, so represents an (unparameterised) trajectory of the dynamical system. A~super-integrable system with the maximal number of functionally independent integrals is called {\em maximally super-integrable}. ``Solving'' the system of equations defining~${\cal S}$, gives the solution, but this cannot in general be determined explicitly. Being only $5$ equation in a~$6$-dimensional space, this solution will depend upon a single additional parameter (as well as the parameters~$c_i$), which will be some function of~$t$, but not necessarily~$t$ itself.

If we start with a separable system of Section~\ref{Sect:separable}, depending upon $3$ arbitrary, single-variable functions, then each additional integral imposes differential constraints on these arbitrary functions. Our maximally super-integrable systems depend on a finite number of arbitrary {\it para\-me\-ters}, whose coefficients are {\it specific} functions (rational in our examples). The set of functions, $\{H,F_i\}_{i=1}^4$, will then generate a non-Abelian Poisson algebra, which may or may not be finite-dimensional.

To simplify all of these calculations, we choose $F_3$, $F_4$ to be a pair of functions whose leading order parts (in momenta) {\it commute}, but allow for the case $\{F_3,F_4\}\neq 0$, in which case
\begin{gather*}
\{F_3,F_4\} = \sum_{i=1}^3 X_i({\boldsymbol{q}}) p_i
\end{gather*}
is a {\it first order} integral.

By choosing the leading order terms of each integral $\{F_i\}_{i=1}^4$ to be just $K^2$, for some ele\-ment~$K$ of the symmetry algebra $\mathfrak{g}$, the automorphisms $\iota_1$ and $\iota_{23}$ (where appropriate) of $\mathfrak{g}$ induce corresponding automorphisms of the Poisson algebra generated by $\{F_i\}_{i=1}^4$. This will be important when deriving the Poisson relations on the full Poisson algebra.

\subsection{Constant curvature case of (\ref{H1234})}\label{sec:eifi}

Here we have the symmetry algebra (\ref{diag-alg}), with Casimir $H_0 = q_1^2\big(p_1^2-p_2^2-p_3^2\big)$. We start with the involutive system given in (\ref{e1e2}), with integrals
\begin{gather*}
H = q_1^2\big(p_1^2-p_2^2-p_3^2\big)+h({\boldsymbol{q}}), \qquad F_1 = e_1^2+g_1({\boldsymbol{q}})= p_2^2+g_1({\boldsymbol{q}}),\\
 F_2 = e_2^2+g_2({\boldsymbol{q}})= p_3^2+g_2({\boldsymbol{q}}),
\end{gather*}
where the potential functions, $h({\boldsymbol{q}})$, $g_1({\boldsymbol{q}})$, $g_2({\boldsymbol{q}})$, are given by (\ref{eesols1}). We then add two further functions
\begin{gather*}
F_3 = f_1^2+g_3({\boldsymbol{q}})\qquad\mbox{and}\qquad F_4 = \frac{1}{4}f_2^2+g_4({\boldsymbol{q}}),
\end{gather*}
where $f_1$, $f_2$ are defined in the list (\ref{diag-alg}). When we impose the conditions $\{H,F_3\}=\{H,F_4\}=0$, it is a simple calculation to derive the following solution:
\begin{gather}
h=q_1^2 \left(\frac{k_1}{q_2^2}+\frac{k_2}{q_3^2}\right), \qquad g_1 = - \frac{k_1}{q_2^2}, \qquad g_2 = - \frac{k_2}{q_3^2}, \nonumber\\
 g_3 = - \frac{k_1 \big(q_2^2+q_3^2-q_1^2\big)^2}{q_2^2},\qquad g_4 = - \frac{k_2 \big(q_2^2+q_3^2-q_1^2\big)^2}{q_3^2}.\label{eeff-pots}
\end{gather}
In this case, we also find that $h_1$ is a first integral.

We therefore have $6$ first integrals $(H,F_1,F_2,F_3,F_4,h_1)$, but the rank of the Jacobian is $5$, so there should be an algebraic relation between them. Nevertheless, we consider these $6$ functions as generators of our Poisson algebra. Under the action of the involutions (automorphisms of the symmetry algebra) $\iota_1$ and $\iota_{23}$, we have
\begin{gather*}
\begin{split}&
\iota_1\colon \ (H,F_1,F_2,F_3,F_4,h_1,k_1,k_2) \mapsto (H,F_3,F_4,F_1,F_2,-h_1,k_1,k_2), \\
& \iota_{23}\colon \ (H,F_1,F_2,F_3,F_4,h_1,k_1,k_2) \mapsto (H,F_2,F_1,F_4,F_3,h_1,k_2,k_1),\end{split}
\end{gather*}
so the entire Poisson algebra should obey such symmetry rules. We can use this in the derivation of the Poisson algebra. For example, if we know the formula for $\{F_1,F_3\}$, then we can use $\iota_{23}$ to deduce the formula for $\{F_2,F_4\}$. Whenever we introduce a new element of our algebra, we should simultaneously introduce any new elements which are derived through the action of these involutions. In this way we add a further~$5$ elements, which, by construction, satisfy $\{H,F_i\}=0$, for all $i$.

Some Poisson relations are very simple to derive
\begin{gather*}
\{F_1,F_2\}=\{F_3,F_4\}=0\qquad \mbox{and}\qquad \{F_i,h_1\}=\lambda_i F_i,\qquad i=1,\dots ,9,
\end{gather*}
where $\lambda = (4,4,-4,-4,0,0,4,0,-4)$.

The Poisson brackets $\{F_1,F_3\}$, $\{F_1,F_4\}$, $\{F_2,F_3\}$, $\{F_2,F_4\},$ are all cubic in momenta and could be {\it linear} combinations of $\{h_1 F_i,h_1 H, h_1^3\}_{i=1}^4$, but this is not the case.
However, note that $\{F_1,F_3\}$ and $\{F_2,F_4\}$ are related through the involutions, as are $\{F_1,F_4\}$ and $\{F_2,F_3\}$. We can define two new {\it quadratic} elements $F_5, F_6$ through the relations
\begin{gather*}
\{F_1,F_3\} = h_1\big(h_1^2-4H-4F_5-4k_1\big),\qquad \{F_2,F_4\} = h_1\big(h_1^2-4H-4F_6-4k_2\big),
\end{gather*}
with $F_5\leftrightarrow F_6$ under $\iota_{23}$. These functions can be written
\begin{gather*}
F_5 = q_3^2F_1+q_1^2F_2+q_3p_3(q_3p_3+2 q_1p_1),\qquad F_6 = q_2^2F_2+q_1^2F_1+q_2p_2(q_2p_2+2 q_1p_1).
\end{gather*}
We define $F_7,F_9$ by the equations
\begin{gather*}
\{F_1,F_6\}= 2 h_1 F_1+4 F_7,\qquad \{F_3,F_6\}= -2 h_1 F_3+4 F_9,
\end{gather*}
related by $F_7\leftrightarrow F_9$ under $\iota_{1}$.

The function $F_8$ is defined by the second of the following equations
\begin{gather*}
\{F_1,F_4\}+\{F_2,F_3\}= 8h_1\left(H+F_5+F_6-\frac{1}{4} h_1^2\right),\qquad \{F_1,F_4\}-\{F_2,F_3\}= 16 F_8,
\end{gather*}
after which, we find that $\{F_5,F_6\}=4 F_8$.

\begin{figure}[hbt]
\unitlength=0.55mm
\centering
\subfigure[Four connected bracket relations.]{
\begin{tikzpicture}[scale=0.6,->,>=stealth',shorten >=1pt,auto,node distance=2.8cm, semithick]
\draw [->] (0.5,0.3) -- (4.5,2);
\draw [->] (0.5,-0.3) -- (4.5,-2);
\draw [->] (5.5,2) -- (9.4,0.2);
\draw [->] (5.5,-2) -- (9.4,-0.2);
\node at (0,0) {$P_{16}$};
\node at (5,2.3) {$P_{36}$};
\node at (5,-2.3) {$P_{25}$};
\node at (10,0) {$P_{45}$};
\node at (2.5,1.6) {$\iota_1$};
\node at (2.5,-1.6) {$\iota_{23}$};
\node at (7.6,1.5) {$\iota_{23}$};
\node at (7.6,-1.5) {$\iota_1$};
\end{tikzpicture}\label{F1636}}
\qquad \subfigure[Two connected bracket relations.]{\begin{picture}(100,50)\thicklines
%==========nodes==============
\put(25,30){\makebox(0,0){$P_{13}$}} \put(75,30){\makebox(0,0){$P_{24}$}}
%=========vectors========================
\put(30,30){\vector(2,0){40}}
%==========labels==============================
\put(50,35){\makebox(0,0){$\iota_{23}$}}
\end{picture}\label{F1324}}
\caption{Bracket relations connected through $\iota_1$ and $\iota_{23}$.} \label{F16361324}
\end{figure}
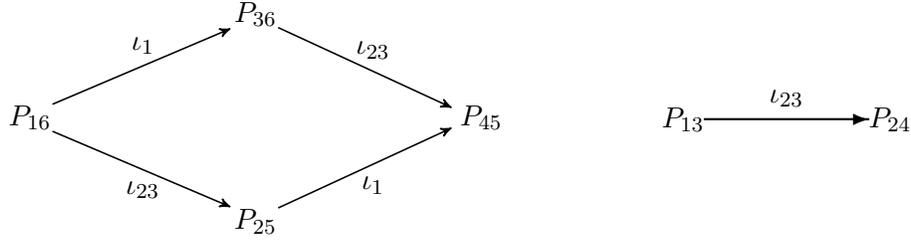

The action of the two involutions is then given by:
\begin{gather*}
\begin{array}{|c||c|c|c|c|c|c|c|c|c|c|c|c|c|}\hline
& H & F_1 & F_2 & F_3 & F_4 & F_5 & F_6 & F_7 & F_8 & F_9 & h_1 & k_1 & k_2 \\ \hline\hline
\iota_1\colon & H & F_3 & F_4 & F_1 & F_2 & F_5 & F_6 & F_9 & F_8 & F_7 & -h_1 & k_1 & k_2\\ \hline
\iota_{23}\colon & H & F_2 & F_1 & F_4 & F_3 & F_6 & F_5 & -F_7 & -F_8 & -F_9 & h_1 & k_2 & k_1\\ \hline
\end{array}
\end{gather*}
The action of $\iota_{23}$ on $\{F_1,F_6\}$ and $\{F_3,F_6\}$ then gives
\begin{gather*}
\{F_2,F_5\}= 2 h_1 F_2-4 F_7,\qquad \{F_4,F_5\}= -2 h_1 F_4-4 F_9.
\end{gather*}
This phenomenon of connecting four different Poisson relations through the involutions is depicted in Fig.~\ref{F1636}, where we define $P_{ij}=\{F_i,F_j\}$ (see Table~\ref{Tab:10Pois_alg}).
Sometimes only {\it two} relations are connected, such as with $P_{13}$ and $P_{24}$ (Fig.~\ref{F1324}), or even just {\it one}, such as with $P_{79}$, because of invariance properties.

\br[commutativity]
The actions of $\iota_1$ and $\iota_{23}$ commute on these functions.
\er

After inputting these known brackets, it is possible to use the Jacobi identity to derive all the others.
The full set of Poisson relations is given in Table~\ref{Tab:10Pois_alg}, with the array $P=(P_{ij})$, using the order $(F_1,F_2,F_3,F_4,F_5,F_6,F_7,F_8,F_9,F_{10}=h_1)$. The lower part of the matrix is given by skew-symmetry. In this context, $H$ is just a~parameter, since it commutes with all $10$ elements. We find
\begin{gather*}
 P_{13} = h_1 \big(h_1^2-4 (H+F_5+k_1)\big), \\
 P_{14}= h_1 \big(4(H+F_5+F_6)-h_1^2\big)+8 F_8, \\
 P_{57}= P_{18} = \frac{1}{2} ((F_1+F_2)\big(4H+4F_5-h_1^2\big)+4 (F_1 F_6+k_1 F_2) ), \\
 P_{19} = 2 (2 (F_5+H) (2 (F_6+H )+F_5 )-F_1 F_4 ) -h_1^2 (3 F_5+2 F_6+4 H-k_2 )+\frac{1}{2}h_1^4 \\
 \hphantom{P_{19} =}{} -4 (k_2 (F_5+H+k_1 )+F_5 k_1 ), \\
 P_{58}= H (4 F_6-h_1^2+4 H)+\frac{1}{2} F_5 \big(8 F_6-h_1^2+12 H\big)+2 F_5^2-2 (k_1+k_2) F_5\\
 \hphantom{P_{58}=}{} +\frac{1}{2} k_2 \big(h_1^2-4 H\big)-2 k_1 k_2, \\
 P_{78} = \frac{1}{4} h_1 (F_1+F_2) \big(h_1^2-4 (H+F_5+F_6)\big).
\end{gather*}
From the list given above, we can derive all except $P_{79}$ by using the involutions (as in Fig.~\ref{F16361324}). The relevant groupings are
\begin{gather*}
 (P_{13},P_{24}),\qquad (P_{14},P_{23}),\qquad (P_{18},P_{28},P_{38},P_{48}),\qquad (P_{19},P_{29},P_{37},P_{47}), \\
 (P_{57},P_{59},P_{67},P_{69}),\qquad (P_{58},P_{68}),\qquad (P_{78},P_{89}).
\end{gather*}
For example, applying $\iota_{23}$ to the formula for $P_{13}$, we get $P_{24}=h_1 (h_1^2-4 (H+F_6+k_2))$. The equality $P_{57}= P_{18}$ follows from the Jacobi identity for the elements $F_1$, $F_5$, $F_6$. The most complicated entry in matrix $P$ is $P_{79}$, which is not obtainable in this way, since it is invariant (up to a sign) under both involutions:
\begin{gather*}
 P_{79} = 2 (F_2F_9-F_4F_7) +\frac{1}{4} h_1 \big(4(H+F_5+F_6)-h_1^2\big)\big(h_1^2-4H-4F_5-2F_6\big) \\
\hphantom{P_{79} =}{} +\frac{1}{2} k_1 h_1 \big(4(H+F_5+F_6)-h_1^2\big)+k_2h_1 \big(4(H+F_5)-h_1^2\big) +4 h_1 k_1 k_2.
\end{gather*}
Under $\iota_1$, $P_{79}\mapsto -P_{79}$, as it should. Under $\iota_{23}$, we should have $P_{79}\mapsto P_{79}$, but, in fact, $P_{79}\mapsto P_{79}+I_{79}$, where
\begin{gather*}
I_{79}= 2 (F_3+F_4)F_7-2(F_1+F_2)F_9+\frac{1}{2}h_1(F_5-F_6)\big(4(H+F_5+F_6-k_1-k_2)-h_1^2\big)\\
\hphantom{I_{79}=}{} +\frac{1}{2}(k_1-k_2)h_1\big(4H-h_1^2\big),
\end{gather*}
which satisfies $I_{79}\mapsto -I_{79}$ under both involutions. However, this does not pose a contradiction, since in the explicit form of the Poisson algebra, $I_{79}=0$.

\begin{table}[h]\centering
\caption{The 10-dimensional Poisson algebra $\{F_i,F_j\}=P_{ij}$.}\label{Tab:10Pois_alg}\vspace{1mm}
{\footnotesize
\begin{gather*}
P=\left(
\begin{matrix}
 0 & 0 & P_{13} & P_{14} & 0 & 4 F_7+2 F_1 h_1 & 2 F_1 F_2 & P_{18} & P_{19} & 4 F_1 \\
 & 0 & P_{23} & P_{24} & 2 F_2 h_1-4 F_7 & 0 & -2 F_1 F_2 & P_{28} & P_{29} & 4 F_2 \\
 & & 0 & 0 & 0 & 4 F_9-2 F_3 h_1 & P_{37} & P_{38} & 2 F_3 F_4 & -4 F_3 \\
 & & & 0 & -4 F_9-2 F_4 h_1 & 0 & P_{47} & P_{48} & -2 F_3 F_4 & -4 F_4 \\
 & & & & 0 & 4 F_8 & P_{57} & P_{58} & P_{59} & 0 \\
 & & & & & 0 & P_{67} & P_{68} & P_{69} & 0 \\
 & & & & & & 0 & P_{78} & P_{79} & 4 F_7 \\
 & & & & & & & 0 & P_{89} & 0 \\
 & & & & & & & & 0 & -4 F_9 \\
 & & & & & & & & & 0
\end{matrix}
\right),
\end{gather*}
}
\end{table}

The functions $(H,F_1,F_2,F_3,F_4)$ are functionally independent integrals of the Hamiltonian system, with Hamiltonian $H$. The functions $(F_5,F_6,F_7,F_8,F_9,h_1)$ must therefore satisfy six functionally independent relations. These can be obtained by looking at the full set of Jacobi identity relations, some of which have non-trivial entries, all of which vanish when the functional forms of $F_i$ are inserted. We label such functions by the Jacobi identity which gave rise to them, so~$J_{ijk}$ is (up to an overall constant multiple) the entry corresponding to $F_i$, $F_j$, $F_k$. These also belong to ``families'', which are related through the action of the involutions. Six such functions are
\begin{gather*}
J_{159} = 2 F_3 F_7-4 F_5 F_8+2 F_1 F_9-4 F_8 H+F_8 h_1^2 +4 k_1 F_8, \\
J_{269} = 2 F_4 F_7-4 F_6 F_8+2 F_2 F_9-4 F_8 H+F_8 h_1^2+4 k_2 F_8, \\
J_{137} = 2 F_7 \big(h_1^2+4 H-4 k_1+4 F_5\big)-8 F_1 F_8\\
\hphantom{J_{137} =}{} + 4h_1 (F_2 H+ F_2 k_1+ F_1 k_2+ (F_1+F_2) F_5)-F_2 h_1^3, \\
J_{139} = 2 F_9 \big(h_1^2+4 H-4 k_1+4 F_5\big)-8 F_3 F_8\\
\hphantom{J_{139} =}{} - 4h_1 (F_4 H+ F_4 k_1+ F_3 k_2+ (F_3+F_4) F_5)+F_4 h_1^3, \\
J_{247} = -2 F_7 \big(h_1^2+4 H-4 k_2+4 F_6\big)+8 F_2 F_8\\
\hphantom{J_{247} =}{} + 4h_1 (F_1 H+ F_1 k_2+ F_2 k_1+ (F_1+F_2) F_6)-F_1 h_1^3, \\
J_{249} = -2 F_9 \big(h_1^2+4 H-4 k_2+4 F_6\big)+8 F_4 F_8\\
\hphantom{J_{249} =}{} - 4h_1 (F_3 H+ F_3 k_2+ F_4 k_1+ (F_3+F_4) F_6)+F_3 h_1^3.
\end{gather*}
These functions satisfy $J_{ijk}=0$, and their Jacobian (with respect to the functions $H,F_i$) has rank $6$, thus giving us the necessary six relations on our algebra.

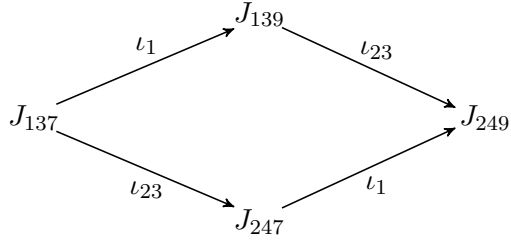
\begin{figure}[hbt]
\unitlength=0.55mm
\centering
\begin{tikzpicture}[scale=0.6,->,>=stealth',shorten >=1pt,auto,node distance=2.8cm, semithick]
\draw [->] (0.5,0.3) -- (4.5,2);
\draw [->] (0.5,-0.3) -- (4.5,-2);
\draw [->] (5.5,2) -- (9.4,0.2);
\draw [->] (5.5,-2) -- (9.4,-0.2);
\node at (0,0) {$J_{137}$};
\node at (5,2.3) {$J_{139}$};
\node at (5,-2.3) {$J_{247}$};
\node at (10,0) {$J_{249}$};
\node at (2.5,1.6) {$\iota_1$};
\node at (2.5,-1.6) {$\iota_{23}$};
\node at (7.6,1.5) {$\iota_{23}$};
\node at (7.6,-1.5) {$\iota_1$};
\end{tikzpicture}
\caption{Constraints connected through $\iota_1$ and $\iota_{23}$.} \label{Jijk}
\end{figure}

\br[action of the involutions] $J_{159}$ and $J_{269}$ are invariant under $\iota_1$ and transform into one-another (up to sign) under $\iota_{23}$. $J_{137}$, $J_{139}$, $J_{247}$ and $J_{249}$ are also connected, as depicted in Fig.~\ref{Jijk}.

The very simple relation
\begin{gather*}
J_{136} = F_2 F_3-F_1 F_4+2 F_8 h_1,
\end{gather*}
also exists and is invariant (up to sign) under the action of both involutions.
\er

\br[comparison with the literature]
The Hamiltonian $H$, with potential $h$ given in~(\ref{eeff-pots}), can be written
\begin{gather*}
H = q_1^2 \left(p_1^2-p_2^2-p_3^2+\frac{k_1}{q_2^2}+\frac{k_2}{q_3^2}\right),
\end{gather*}
which is in the form of a St\"ackel transform to a flat metric, so can be compared with the classification given in~\cite{17-3}. We can compare with the list of non-degenerate potentials (albeit in the quantum case with $H_0$ replaced by a Laplacian) given in~\cite[Section~5]{17-3}. Allowing for the fact that this classification is for the Euclidean case and involves some complex coordinates, the above potential can be considered as a reduction of the 5 parameter potential~$V_{[2,1,1,1]}$ of~\cite{17-3}; it is, in fact a reduced case, with our parameters $(k_1,k_2)$ corresponding to their $(a_2,a_3)$.
\er

\subsection{The flat case of (\ref{Hija2=0})}

The simplest flat case, with $H$ given by (\ref{H0flat+h}), has integrals of the form
\begin{gather*}
F_1=e_1^2+g_1,\qquad F_2=4e_2^2+g_2,\qquad F_3=f_1^2+g_3,\qquad F_4=f_2^2+g_4,
\end{gather*}
since this is invariant under the action of the involution $\iota_1$, so the Poisson algebra possesses this automorphism. This is the only case we consider here.

The functions $\varphi_1$, $\varphi_2$ of Section \ref{sec:flat-e1e2} are constrained by the additional integrals, giving
\begin{gather}
h=\frac{k_1 (q_1-q_3 )^2}{q_2^2}+\frac{2 k_2 q_1}{q_1-q_3}, \qquad g_1 = - \frac{k_1}{q_2^2}, \qquad g_2 = - \frac{k_2}{ (q_1-q_3 )^2}, \nonumber\\
 g_3 = - \frac{k_1 \big(q_1^2-q_2^2-q_3^2\big)^2}{q_2^2},\qquad g_4 = - \frac{k_2 \big(q_1^2-q_2^2-q_3^2\big)^2}{ (q_1-q_3 )^2}. \label{flat-eeff-pots}
\end{gather}
In this case, we also find that $h_1$ is a first integral.

We therefore have $6$ first integrals $(H,F_1,F_2,F_3,F_4,h_1)$, but the rank of the Jacobian is $5$, so there should be an algebraic relation between them. Nevertheless, we consider these $6$ functions as generators of our Poisson algebra. Under the action of the involution (automorphism of the symmetry algebra) $\iota_1$, we have
\begin{gather*}
 \iota_1\colon \ (H,F_1,F_2,F_3,F_4,h_1,k_1,k_2) \mapsto (H,F_3,F_4,F_1,F_2,-h_1,k_1,k_2),
\end{gather*}
so the entire Poisson algebra should obey such symmetry rules, induced by the definitions below.

We have $\{F_1,F_2\}=\{F_3,F_4\}=0$ and the cubic expression $\{F_1,F_3\}=4h_1F_5$ {\em factorises}, giving us a new {\em quadratic} integral $F_5$, which is invariant under the action of $\iota_1$. Two new {\em cubic} integrals (related through $\iota_1$) are defined by
\begin{gather*}
\{F_1,F_4\}=8F_6 \qquad\mbox{and}\qquad \{F_2,F_3\}=8F_7,
\end{gather*}
whilst $\{F_2,F_4\}$ is just {\em linear}: $\{F_2,F_4\}=-8k_2 h_1$. The first four brackets with $F_5$ are
\begin{gather*}
\{F_1,F_5\}= 2 h_1 F_1, \qquad \{F_2,F_5\}=4F_8, \qquad \{F_3,F_5\}= -2 h_1 F_3, \qquad \{F_4,F_5\}=4F_9,
\end{gather*}
giving us two new {\em cubic} integrals. We find the following factorisation:
\begin{gather*}
F_6+F_7 = h_1 F_{10},
\end{gather*}
which defines another {\em quadratic} integral. All remaining brackets can be determined in terms of the above $F_i$ and $h_1$. In these relations, $H$ acts as a parameter, since it (by definition) commutes with {\em all} $F_i$. The action of the involution $\iota_1$ is then given by:
\begin{gather*}
\begin{array}{|c||c|c|c|c|c|c|c|c|c|c|c|c|}\hline
& H & F_1 & F_2 & F_3 & F_4 & F_5 & F_6 & F_7 & F_8 & F_9 & F_{10} & h_1 \\ \hline\hline
\iota_1\colon & H & F_3 & F_4 & F_1 & F_2 & F_5 & -F_7 & -F_6 & F_9 & F_8 & F_{10} & -h_1\\ \hline
\end{array}
\end{gather*}
The integral $h_1$ acts diagonally:
\begin{gather*}
 \{F_i,h_1\}=\lambda_i F_i,\qquad i=1,\dots ,10,\qquad \mbox{where}\qquad \lambda = (4,4,-4,-4,0,0,0,4,-4,0).
\end{gather*}
Defining $P_{ij}=\{F_i,F_j\}$, the remaining independent entries in the Poisson matrix are
\begin{gather*}
 P_{16}=2F_1(H+3 F_{10}-k_2)+4 k_1F_2,\qquad P_{17}=2 (F_2F_5+F_8h_1),\qquad P_{18}=2 F_1F_2, \\
 P_{19}= -4F_5(H+2F_{10}-k_2)-2F_1F_4-8k_1F_{10},\qquad P_{110}=4F_8,\qquad P_{26}=-4 k_2 F_1,\\
 P_{27}=4(F_2F_{10}+k_2F_1), \qquad P_{28}=0,\qquad
 P_{29} =4F_{10}(k_2-H-F_{10})+8k_2(F_5+2k_1),\\ P_{210} = 0, \qquad P_{56} = 2(F_5(H+2 F_{10}-k_2)-F_6h_1+2k_1F_{10}),\\
 P_{58} = F_2F_5-F_8h_1-2k_1F_2-F_1(H+3 F_{10}-k_2),\qquad P_{510} = 2 (F_7-F_6),\\
 P_{67} = F_4 F_8-F_2 F_9 +h_1 (F_{10} (H-k_2)-2 k_2 \big(F_5+2 k_1)+F_{10}^2\big),\\ P_{68} = 2 (F_2 F_6+F_8 (-2 F_{10}-H+k_2)), \qquad
 P_{69} = 2 F_9 (2 F_{10}+H-k_2)-2 F_4 F_6,\\ P_{610} = 2 \big(F_{10} (H-k_2)-2 k_2 (F_5+2 k_1)+F_{10}^2\big),\\
 P_{89} = -2 h_1 \big(F_{10} H-F_{10} k_2+2 F_{10}^2-4 k_1 k_2\big),\qquad P_{810} = 4 F_1 k_2+2 F_2 F_{10}.
\end{gather*}
The remaining entries can be obtained by using the involution $\iota_1$.

The functions $(H,F_1,F_2,F_3,F_4)$ are functionally independent integrals of the Hamiltonian system, with Hamiltonian $H$. The functions $(F_5,F_6,F_7,F_8,F_9,F_{10},h_1)$ must therefore satisfy seven functionally independent relations. Again, these can be obtained by looking at the full set of Jacobi identity relations, some of which have non-trivial entries, all of which vanish when the functional forms of $F_i$ are inserted. We label such functions by the Jacobi identity which gave rise to them, so $J_{ijk}$ is (up to an overall constant multiple) the entry corresponding to~$F_i$,~$F_j$,~$F_k$. These also arise in ``pairs'', which are related through the action of the involution $\iota_1$. The first, in fact, comes from the {\em definition} of~$F_{10}$ (which satisfies $I_{10}\mapsto -I_{10}$ under the action of $\iota_1$):
\begin{gather*}
 I_{10} = F_6+F_7-h_1 F_{10} ,\qquad J_{1310} = F_2 F_3-F_1 F_4+(F_6-F_7) h_1 ,\\
 J_{267} = F_1 (F_{10}+H-k_2)+F_2 (F_5+2 k_1)-F_8 h_1 , \\ J_{467} = F_3 (F_{10}+H-k_2)+F_4 (F_5+2 k_1)+F_9 h_1 ,\\
 J_{146} = F_4 F_8 -F_6 (F_{10}+H-k_2)+ k_2h_1 (F_5+2 k_1) ,\\ J_{237} = F_2 F_9+F_7 (F_{10}+H-k_2)- k_2h_1 (F_5+2 k_1) ,\\
 J_{1610} = F_2 F_6 -F_8 (F_{10}+H-k_2)+ k_2 h_1 F_1.
\end{gather*}

\subsubsection{Flat coordinates}

We saw that the flat coordinates (\ref{flat-coord}) reduce this $H_0$ to the form~(\ref{H0-flat-coord}). In these coordinates, the first~5 (functionally independent) integrals of our Poisson algebra take the form
\begin{gather*}
 H=2P_1P_3-P_2^2+\frac{k_1}{Q_2^2}+k_2\big(1+Q_2^2-2 Q_1Q_3\big), \\
 F_1=4\left(Q_2^2P_1^2+Q_3^2\left(P_2^2-\frac{k_1}{Q_2^2}\right)+2 Q_2Q_3P_1P_2\right), \\
F_2=4\big(P_1^2-k_2Q_3^2\big),\qquad F_3=(Q_1P_2+Q_2P_3)^2-k_1 \frac{Q_1^2}{Q_2^2} , \qquad F_4 = P_3^2-k_2Q_1^2 .
\end{gather*}
The functions $F_5,\dots ,F_{10}$ can similarly be found and since the transformation is canonical, the Poisson relations do not change.

In these coordinates, the involution $\iota_1$ is generated by $S=-\big(2 q_3P_1+q_2 P_2+\frac{1}{2}q_1P_3\big)$.

\br[comparison with the literature]
The Hamiltonian $H$, with potential $h$ given in~(\ref{flat-eeff-pots}), can be written
\begin{gather*}
H = (q_1-q_3)^2 \left(p_1^2-p_2^2-p_3^2+\frac{k_1}{q_2^2}+\frac{2k_2 q_1}{(q_1-q_3)^3}\right),
\end{gather*}
which is in the form of a St\"ackel transform to a flat metric, so, again, can be compared with the classification given in \cite{17-3}. We can compare with the list of non-degenerate potentials (albeit in the quantum case with $H_0$ replaced by a Laplacian) given in of \cite[Section~5]{17-3}. Allowing for the fact that this classification is for the Euclidean case and involves some complex coordinates, the above potential can be considered as a reduction of the 5 parameter potential $V_{[2,2,1]}$ of~\cite{17-3}; it is, in fact a reduced case, with our parameters $(k_1,k_2)$ corresponding to their $(a_1,a_3)$.

In the flat coordinates (diagonalised), it again corresponds to $V_{[2,2,1]}$, but now with $(k_1,k_2)$ corresponding to their $(a_1,a_4)$.
\er

\subsection{The conformally flat case (\ref{I2+h})}\label{conflat-super}

In this section we consider the involutive system (\ref{I2+h}), with $K=e_1$, giving the potential functions~(\ref{K=e1}). To simplify the calculations of this section, we choose the specific metric coefficient $\varphi(r_0)=r_0$, leading to
\begin{gather*}%\label{Hcon+h1}
H = H_0 + h({\boldsymbol{q}}), \qquad\mbox{where}\qquad H_0= q_1 q_3 \big(p_1^2-p_2^2-p_3^2\big),
\end{gather*}
whose Ricci tensor is non-constant: $R=\frac{5\big(q_1^2-q_3^2\big)}{2 q_1q_3}$. In addition to $F_1$, $F_2$, given in~(\ref{I2+h}) and~(\ref{K=e1}), we require that the quadratic function
\begin{gather*}
F_3=\frac{1}{4} h_1^2+g_3({\boldsymbol{q}}),
\end{gather*}
satisfies $\{F_3,H\}=0$, which restricts the component functions of $h({\boldsymbol{q}})$, as well as determining~$g_3$, but leaves $\varphi_1\big(\frac{q_3}{q_1}\big)$ {\em arbitrary}. Specifically, we find
\begin{gather*}
\varphi_2\big(q_3^2-q_1^2\big)=k_0+k_1\big(q_3^2-q_1^2\big)+k_2 \big(q_3^2-q_1^2\big)^2 ,\qquad \varphi_3(q_2)=k_2 q_2^2-\frac{k_3}{2 q_2^2},\\
g_3 = -\big(q_1^2-q_2^2-q_3^2\big)\big(k_1-k_2\big(q_1^2-q_2^2-q_3^2\big)\big).
\end{gather*}
Setting $\varphi_1\big(\frac{q_3}{q_1}\big)=0$ (since it effectively an additive constant), we therefore have
\begin{gather*}
H = q_1q_3\left(p_1^2-p_2^2-p_3^2 +\frac{k_0}{q_1^2-q_3^2} - k_1+k_2 \big(q_1^2-q_2^2-q_3^2\big)+\frac{k_3}{2q_2^2}\right),\\
F_1 = {\cal C}_1 -k_1 \big(q_1^2-q_3^2\big)+k_2 \big(q_1^2-q_3^2\big)\big(q_1^2-q_2^2-q_3^2\big)+\frac{k_3\big(q_1^2-q_3^2\big)}{2q_2^2},\\
F_2 = e_1^2 +k_2 q_2^2-\frac{k_3}{2q_2^2}, \qquad F_3 = \frac{1}{4} h_1^2- \big(q_1^2-q_2^2-q_3^2\big)\big(k_1-k_2(q_1^2-q_2^2-q_3^2\big)\big) .
\end{gather*}
The algebra of these three integrals is easily calculated:
\begin{gather*}
\{F_1,F_2\}=0, \qquad \{F_1,F_3\}=0, \qquad \{F_2,F_3\} = 2F_{4},
\end{gather*}
where $F_{4}$ is a cubic expression
\begin{gather*}
F_{4} = F_2 h_1-2 k_1 q_2 p_2 +4 k_2 q_2 \big(q_2(q_1p_1+q_3p_3)+\big(q_1^2-q_3^2\big)p_2\big),
\end{gather*}
which cannot be written as a polynomial in $H$ and its integrals, but does satisfy the algebraic relation
\begin{gather*}
F_{4}^2=4 F_2^2F_3+4k_1F_2(F_3-F_1)-4k_2 (F_3-F_1)^2+4k_2k_3 (F_3+F_1)\\
\hphantom{F_{4}^2=}{} +2k_1k_3F_2+k_3(2k_1^2-k_2k_3).
\end{gather*}
We can use this to derive the formulae for $\{F_i,F_4\}$: $\{F_1,F_4\}=0$ and
\begin{gather*}
\{F_2,F_4\}=4F_2^2+4k_1F_2-8k_2(F_3-F_1)+4k_2k_3,\\ \{F_3,F_4\}=-8F_2F_3-4k_1(F_3-F_1)-2k_1k_3.
\end{gather*}

\br
We have 4 functionally independent integrals, $H$, $F_1$, $F_2$, $F_3$, which form a closed algebra (with the inclusion of the functionally {\em dependent}~$F_4$). In $3$-dimensions, this system is super-integrable, but {\em not} maximally. Since we only have a $3$-dimensional symmetry algebra $\mathfrak{g}_1$, we cannot build any further integrals out of this. However, since the metric is not constant curvature, it is possible that other integrals exist, which are not built in this way.
\er

\section{Conclusions}

We started this paper by considering a specific $3$-dimensional realisation of~$\mathfrak{sl}(2)$ (our algebra~$\mathfrak{g}_1$) and showed how to embed this into some 6- and 10-dimensional Lie algebras, with very specific structure. The Casimir functions (\ref{Hg221case3}) and (\ref{Hflat}) of the $6$-dimensional algebras, represent the kinetic energy on manifolds with these symmetries, and the $10$-dimensional algebra gave us the corresponding {\em conformal algebra}. This was used in Sections~\ref{Sect:separable} and~\ref{Sect:super}, where we considered some specific diagonal examples of these Casimir functions.

Indeed, the main aim of this paper was to build super-integrable systems (and the associated non-Abelian Poisson algebras) with a given kinetic energy, which itself has a high degree of symmetry. The approach is most suited to kinetic energies derived from constant curvature (including flat) manifolds, as mainly considered here. However, we saw that it is possible to apply the method when the isometry algebra is smaller, but, in this case, it may not be possible to build enough {\em independent} integrals for the system to be {\em maximally} super-integrable.

We just considered $3$-dimensional manifolds in this paper. We used the structure of the symmetry algebra to construct involutive triples, giving us separable systems. We further used the structure of the symmetry algebra to select the leading parts of additional integrals to obtain super-integrable restrictions. We selected our integrals in such a way that the automorphisms of the symmetry algebra (realised as canonical transformations) could be extended to act simply on the non-Abelian Poisson algebra, which enabled us to find a {\em finite closure} of the Poisson algebra.

We did not consider the classification of super-integrable systems, such as can be found in the literature (see, for example,~\cite{17-3,07-9,13-2}), but our approach can be applied to any kinetic energy associated with a~flat or constant curvature metric, which includes most physical systems.

In this paper we made several choices which simplified our calculations, leaving us with a~number of open problems.

Not all realisations of $\mathfrak{sl}(2)$ are equivalent. Whilst $2$-dimensional realisations were classified by Lie (see~\cite[Section~2]{96-6}), no such classification exists for the $3$-dimensional case. Within our general~$\mathfrak{g}_1$, with~$f_1$ given by~(\ref{general-f1}), there are at least {\em two} equivalence classes (corresponding to degenerate and non-degenerate Casimir functions), but we don't have a full classification of inequivalent cases. Given such a~choice of~$\mathfrak{g}_1$, the constructions of Sections~\ref{hw-g1},~\ref{Sect:PoisAlg} and~\ref{Sect:conformal} could be carried out: the general problem is to find all 6- and 10-dimensional, nontrivial extensions of~$\mathfrak{g}_1$, which are then to be used in the construction of separable and super-integrable systems.

We saw in Section~\ref{Sect:conformal} that the 10-dimensional algebra contains several subalgebras, whose Casimir functions correspond to conformally equivalent metrics. The (infinitesimal) conformal factors formed another representation space of the algebra~$\mathfrak{g}_1$. We don't yet have the full classification of subalgebras of our $10$-dimensional conformal algebra. For any subalgebra, the Casimir function, representing the corresponding kinetic energy, could be used in the context of our analysis of Sections~\ref{Sect:separable} and~\ref{Sect:super}.

In Section~\ref{Sect:separable} we made the {\em simplest choice} of involutive triple. More general quadratic forms would clearly lead to more complicated calculations, but could lead to some interesting examples.

In Section~\ref{Sect:super} we chose additional integrals that would minimise the complexity of our calculations. Clearly there are many different choices which could lead to interesting systems and corresponding Poisson algebras. Even for the simple choices we made, the Poisson algebras were very complicated and we have little understanding of the general structure. The Poisson algebras satisfy many polynomial constraints, which can be used to simplify some of the Poisson relations, but what is the minimal set of generators of these constraints?

The current paper is about classical Poisson algebras, but a similar analysis can be carried out for the quantum case, where super-integrability should allow us to construct explicit eigenfunctions, as was found in $2$-dimensions \cite{f07-1}.

\subsection*{Acknowledgements}

This work was carried out while QH was visiting Leeds for one year, funded by the China Scholarship Council. QH would like to thank the School of Mathematics, University of Leeds, for their hospitality. Significant improvements were made after the comments of both referees and an editor. We thank them for their input.

\pdfbookmark[1]{References}{ref}
\LastPageEnding

\end{document}